\documentclass{article}
\usepackage{setspace} 
\usepackage[margin=1in]{geometry}
\usepackage{wrapfig}
\usepackage{subfig}
\usepackage{graphicx}
\usepackage{float}
\usepackage{bbm}
\usepackage[title]{appendix}
\usepackage{mathtools}     
\usepackage{multirow}
\usepackage{natbib}
\usepackage{amsmath, amssymb, enumerate, amsthm}
\usepackage{amsfonts}
\usepackage{bigints}
\usepackage{booktabs}  
\usepackage{mathrsfs}
\usepackage{scalerel}
\usepackage{authblk}
\usepackage{xcolor}
\usepackage[colorlinks=true,
            pdfpagemode=UseNone,
            urlcolor=blue,
            citecolor=blue,
            linkcolor=blue,
            bookmarks=true,
            backref=page
            ]{hyperref}

\newcommand{\rdd}{\mathbb{R}^{d}}
\newcommand{\re}{\mathbb{R}}

\newcommand{\kerr}{\mathcal{K}}
\newcommand{\kerrO}{\mathscr{K}}
\newcommand{\ltwo}{L^2}
\newcommand{\norm}[1]{\left\lVert#1\right\rVert}
\newcommand{\eqd}{\stackrel{\text{d}}{=}}

\newcommand{\sumn}{\sum_{i=1}^n}

\newcommand{\Prr}[1]{\Pr\left(#1\right)}

\newcommand{\cond}{\stackrel{\text{d}}{\rightarrow}}

\newcommand{\Dd}{{\rm D}}

\newcommand{\E}[2]{{\rm {E}}_{#1}\left[#2\right]}
\newcommand{\Ee}[2]{{\rm {E}}_{#1}#2}

\newcommand{\Eeee}{{\rm E}}
\newcommand{\bfr}{\mathbf{r}}
\newcommand{\bfk}{\mathbf{k}}

\DeclareMathOperator*{\argmax}{ {\rm argmax}}

\DeclareMathOperator*{\argmin}{ {\rm argmin}}

\DeclarePairedDelimiter\floor{\lfloor}{\rfloor}
\DeclarePairedDelimiter\ip{\langle}{\rangle}

\DeclareMathOperator{\med}{Med}

\newcommand{\bbN}{\mathbb{N}}
\newcommand{\cM}{\mathcal{M}}

\newcommand{\cG}{\mathcal{G}}
\newcommand{\cW}{\mathcal{W}}

\newcommand{\cF}{\sF}

\newcommand{\scA}{\mathscr{A}}

\newcommand{\sF}{\mathscr{F}}

\newcommand{\sB}{\mathscr{B}}

\newcommand{\VC}{\operatorname{VC}}

\DeclareMathOperator{\ID}{ID}

\DeclareMathOperator{\HD}{HD}

\DeclareMathOperator{\RP}{RPD}

\DeclareMathOperator{\MFHD}{MFHD}
\DeclareMathOperator{\SD}{SD}
\DeclareMathOperator{\SW}{SW}

\numberwithin{equation}{section}

\newtheorem{theorem}{Theorem}[section]
\newtheorem{remark}{Remark}[section]
\newtheorem{corollary}{Corollary}[section]
\newtheorem{definition}{Definition}[section]

\newtheorem{lemma}{Lemma}[section]

\newtheorem{condition}{Condition}
\providecommand{\keywords}[1]
{
  \small	
    \begin{center}\textbf{\textit{Keywords---}} #1\end{center}
}
\title{Robust changepoint detection in the variability of multivariate functional data}
\author[1]{Kelly Ramsay}
\author[2]{Shoja'eddin Chenouri}
\affil[1]{Department of Mathematics and Statistics, Ross South Building, Office 511A, York University, 4700 Keele St, Toronto, Ontario, Canada. \textit{kramsay2@yorku.ca}}
\affil[2]{Department of Statistics and Actuarial Science, University of Waterloo, 200 University Ave W, Waterloo, Ontario, Canada}
\date{}

\begin{document}
\maketitle

\begin{abstract}
We consider the problem of robustly detecting changepoints in the variability of a sequence of independent multivariate functions. 
We develop a novel changepoint procedure, called the functional Kruskal--Wallis for covariance (FKWC) changepoint procedure, based on rank statistics and multivariate functional data depth. 
The FKWC changepoint procedure allows the user to test for at most one changepoint (AMOC) or an epidemic period, or to estimate the number and locations of an unknown amount of changepoints in the data. 
We show that when the ``signal-to-noise'' ratio is bounded below, the changepoint estimates produced by the FKWC procedure attain the minimax localization rate for detecting general changes in distribution in the univariate setting (Theorem~\ref{thm::main-result}). 
We also provide the behavior of the proposed test statistics for the AMOC and epidemic setting under the null hypothesis (Theorem~\ref{thm::null}) and, as a simple consequence of our main result, these tests are consistent (Corollary~\ref{cor::ell_known}). 
In simulation, we show that our method is particularly robust when compared to similar changepoint methods. 
We present an application of the FKWC procedure to intraday asset returns and f-MRI scans. 
As a by-product of Theorem~\ref{thm::main-result}, we provide a concentration result for integrated functional depth functions (Lemma~\ref{lem::cb}), which may be of general interest.
\end{abstract}
\keywords{Depth function, Multiple changepoint, Covariance operator, Rank statistics}
\section{Introduction}

Motivated by \citet{Stoehr2019}, who expressed the need for a robust method for detecting changepoints in the covariance operator of f-MRI data, we study the problem of robustly and nonparametrically detecting changepoints in the variability of a sequence of multivariate functional data. 
Detecting the presence and location of changepoints in the covariance operator of a sequence of observed functions has received some recent interest in the statistics literature, see, e.g., \citep{Jaruskova2013, Sharipov2019, Jiao2020, Dette2020a, Harris2021}. 
However, most previous works have not considered the robustness of their procedure. 
For example, many previous works require fourth moment assumptions \citep{Stoehr2019,Dette2020a, Sharipov2019}, are based on CUSUM statistics and/or long-run covariance estimators which are not robust \citep{Dette2020a, Sharipov2019, Jiao2020} and/or rely on bootstrapping or other data-driven methods to estimate the null distribution of the test statistic, which are vulnerable to corruption \citep{Sharipov2019, Dette2020a,Jiao2020}. 
Recently, \cite{Harris2021} began the study of robust changepoint detection in the mean and/or covariance operator of an observed sequence of functions. 
They used simulation to test the performance of a fused Lasso and a CUSUM statistic for robustly detecting multiple changepoints in the mean and/or covariance operator of an observed sequence of functions. 

We continue the study of robust changepoint detection in multivariate functional data by providing a comprehensive changepoint detection procedure which detects changes in the variability of such data. 
In particular, we leverage the recent results of \citet{Ramsay2023b}, who show that a large class of differences in the covariance kernel between samples imply differences in the median of multivariate functional data depth values between samples. 
We call this class of differences ``differences in variability''.  
Motivated by the results of \citet{Ramsay2023b}, we study depth-based ranking methods in the setting of changepoint detection in variability. 

Our contributions are as follows: We introduce a nonparametric, robust procedure for detecting changes in the variability of multivariate functional data. We may call this procedure the Functional Kruskal--Wallis Covariance operator changepoint procedure, following the naming convention of \cite{Ramsay2023b}. The FKWC changepoint procedure includes a hypothesis test for the presence of at most one changepoint (AMOC), a hypothesis test for the presence of an ``epidemic period'' and an algorithm to estimate the locations of multiple changepoints when the number of changepoints is not known. 

Our main result says that when the ``signal-to-noise'' ratio is bounded below, the estimated changepoints resulting from the FKWC procedure attain the minimax localization rate for detecting general changes in distribution in the univariate setting \citep{Padilla2021} (Theorem~\ref{thm::main-result}). 
We also derive the asymptotic distribution of the proposed test statistics under the null hypothesis for the AMOC and epidemic setting (Theorem~\ref{thm::null}) and, as a simple consequence of our main result, these tests are consistent (Corollary~\ref{cor::ell_known}). 
As a by-product of Theorem~\ref{thm::main-result}, we prove a concentration result for integrated functional depth functions (Lemma~\ref{lem::cb}), which may be of general interest. 
In addition, another by-product of Theorem~\ref{thm::main-result} is an improved rate of convergence for the multivariate changepoint estimator given by \cite{RAMSAY2023}, see Remark \ref{rem::mv_est}. 
A simulation study and a real data analysis are also included, which highlight the robustness of the proposed procedure (see Section~\ref{sec::sim} and Section \ref{sec::DA}).

The rest of the paper is organized as follows. 
Section \ref{sec::methods} introduces the assumed changepoint model and outlines the FKWC changepoint procedure. 
Section \ref{sec::theory} presents our main theoretical results. 
Section \ref{sec::sim} presents a simulation study, where we compare the FKWC methods to those of \citep{Sharipov2019, Dette2020a, Harris2021}. 
Section \ref{sec::DA} presents applications of our methods to intraday stock returns and resting state f-MRI scans. 
\section{Methodology}\label{sec::methods}
\subsection{Preliminary assumptions}
We suppose that we have observed a sequence of mutually independent multivariate random functions $X_1,\ldots,X_n$ defined on some probability space $(\Omega',\scA',P)$. 
For each $\omega\in\Omega'$, each $X_i\coloneqq X_{i,\omega}$ is a map from $[0,1]^d$ to $\re^p$. 
We assume that for some integers $k_0=0<k_1<\dots<k_\ell<k_{\ell+1}=n$, it holds that $X_{k_{i-1}+1},\dots, X_{k_{i}}$ have a common law $\nu^i$. 
For $m\in\bbN$, let $[m]=\{1,\ldots,m\}$. 
It is assumed that for all $i\in[\ell]$, we have that $\nu^{i}\neq \nu^{i+1}$. Furthermore, we assume that $\nu^i\neq \nu^j$ in such a way so that the shape and/or magnitude of functions drawn from $\nu^i$ differ from that of $\nu^j$, on average. 
We call this a change in variability.\footnote{Formally, our procedure can detect changes in the sequence of (population) functional depth values, to be defined in Section \ref{sec::depth}.}
Define the covariance kernel and covariance operator of $\nu^i$, respectively, as follows
$$
\kerr_i(s,t)=\E{}{X_{k_i}(s)X_{k_i}^\top(t)} -\E{}{X_{k_i}(s)}\E{}{X_{k_i}(t)}^\top\qquad\text{and}\qquad(\kerrO_i f)(t):=\int_{[0,1]^d} f(s) \kerr_i(s,t) d s.
$$
Functional depth values are designed to differentiate between functions with differing shape and/or magnitude \citep{Sun2011, DAI2020, Ramsay2023b}, and so it is natural to use them in this setting. 
We will consider two cases, the first of which is the case when $\ell$ is known. Here, the goal is to test if there are changepoints in the sequence of data. 
If the test is significant, the goal is then to estimate each $k_i$, the locations of the changepoints. 
The second case is where $\ell$ is unknown, and the goal is to estimate each $k_i$ and $\ell$, both the locations and number of changepoints.

It is necessary to clarify what is meant by the ``function'' in the preceding paragraph. 
We assume that the sequence of observations satisfies the following condition.
\begin{condition}\label{cond::contin}
For $d,p\in\bbN$ and all $i,j\in[n]$ with $i\neq j$ it holds that
\begin{itemize}
    \item $\E{}{X_i}=0$ and $X_i$ is independent of $X_j$,
    \item $X_i$ is a continuous function,
    \item $X_i$ is a mean square continuous stochastic process,
    \item each component of $X_i$ is differentiable on $(0,1)^{d}$ and each of its partial derivatives are mean square continuous stochastic processes and continuous functions.
\end{itemize}
\end{condition}
Let $\ltwo\coloneqq \ltwo[0,1]^d$ be the space of square integrable functions over $[0,1]^d$ and let $(\ltwo)^p$ denote the $pth$ Cartesian product of $\ltwo$ with itself. 
Condition~\ref{cond::contin} implies that $X_i\in (\ltwo)^p$. 
In addition, Condition~\ref{cond::contin} ensures that we can view the observed functions and their derivatives as both stochastic processes over $[0,1]^d$ and random elements drawn from some probability measure over $(\ltwo)^p$, see \citep{hsing_eubank_2015} for more details on representations of functional observations. 
When $p>1$, this constitutes the case known as multivariate functional data and when $p=1$ this constitutes the case known as functional data. 
Note that the assumption $\E{}{X_i}=0$ is not particularly restrictive. 
When the observed functions do not have zero mean, they can be centered by a robust measure of location to satisfy the requirement. 
In addition, if one suspects changepoints in the mean of the sequence, a robust changepoint algorithm for detecting such changepoints can be run first. 
One can then center the data within each segment. 


\subsection{Multivariate functional depth functions}\label{sec::depth}
We now cover the essential background on multivariate functional depth functions, on which our proposed methodology relies. 
Denote the empirical measure of the observed functions by $\hat{\nu}$ and, given a space, $\mathscr{S},$ let $\cM_1(\mathscr{S})$ denote the set of probability measures on $\mathscr{S}$. 
As the name suggests, a multivariate functional depth function measures the ``depth'' of an observed function with respect to a given probability measure on $(\ltwo)^p$, i.e., $\ID\colon (\ltwo)^p\times \cM_1((\ltwo)^p)\rightarrow[0,1]$. 
Therefore, the sample depth value of an observed function $X_i$, i.e., $\ID(X_i,\hat{\nu})$, describes the centrality of $X_i$ with respect to $\hat{\nu}$. 
An observation will have high depth (with respect to the sample $\hat\nu$) when it is nested inside the sample and is similar in shape to the sample. 
This fact makes depth a natural choice when detecting changes in variability.

There are many definitions of multivariate functional depth. 
We focus on a general class of integrated depths, or integrated dual depths \citep{Cuevas2009}, which we define below. 
Let $(\Omega,\scA, Q)$ be a probability space, where $(\Omega,\scA)$ is a separable measure space. 
Now, let $g\colon (\ltwo)^p\times\Omega  \to\re^p$ be a measurable function. 
Next, for $u\in\Omega$, define $\nu_u$ to be the law of $g(X,u)$ where $X\sim \nu$. 
Lastly, for a given multivariate depth function $\Dd\colon\re^p\times \cM_1(\re^p)\to [0,1]$, (which is assumed to be a Borel function) define the integrated multivariate functional depth as 
\begin{equation}\label{eqn::integrated_depth}
    \ID(x,\nu)=\int_{\Omega} \Dd(g(x,u),\nu_u) dQ(u).
\end{equation}
Typically in practice, a finite number, say $M$, values from $\Omega$ are selected (either deterministically or randomly), and we use the following approximation
\begin{equation*}
    \ID(x,\hat\nu)\approx\frac{1}{M}\sum_{i=1}^M \Dd(g(x,u_i),\hat\nu_{u_i}).
\end{equation*}
Two popular notions of multivariate functional depth can be recovered from the above definition. 
Namely, the multivariate functional halfspace depth \citep{Fraiman2001, slaets11, Claeskens2014} and the random projection depth \citep{Cuevas2007}. 
The multivariate functional halfspace depth, denoted $\MFHD$, can be recovered by taking $\Omega=[0,1]^d$, $g(x,u)=x(u)$, $Q$ equal to the Lebesgue measure on $[0,1]^d$ and $\Dd=\HD$, the multivariate halfspace depth. 
Let $S$ be the unit sphere in $\ltwo$ and, for $x,y\in L^2$, let $\ip{x,y}=\int_{[0,1]^d}x(t)y(s)dtds$. 
The random projection depth, denoted $\RP$, can be recovered by taking $\Omega=S$, $Q\in \cM_1(S)$, $g(x,u)=(\ip{x_1,u},\ldots,\ip{x_p,u})$ and $\Dd$ to be any multivariate depth. 
Explicit definitions for these instances of the integrated depth can be seen in Appendix \ref{app::depth}. 
Our theoretical results, see Section \ref{sec::theory}, cover the general class of integrated depths given by \eqref{eqn::integrated_depth}, provided there are no issues with measurability. 

\subsection{Changepoint methodology}\label{sec::meth}

The backbone of our changepoint methodology is a form of \emph{sample depth-based ranks}, which incorporate the derivatives of the observed functions. 
For a given function, $x\in(\ltwo)^p$, let $x_k$ be the $k$th component of $x$ and suppose that $x_k$ is differentiable on $(0,1)^d$ for all $k\in[p]$. 
Denote the vector of partial derivatives of each component $k$ of $x$ by $\nabla x_{k}$ and let $h(x)=(x,\nabla x_{1},\ldots, \nabla x_{k})$. 
That is $h(x)$ is the $p(d+1)$-dimensional vector of $x$ and each of its $p$-components' partial derivatives.

Next, let $h\circ\hat\nu$ be the empirical measure of $h(X_1),\ldots,h(X_n)$. 
For a given integrated depth function, we will consider the depth values $\ID(h(X_i), h\circ\hat\nu)$. 
Now, given that depth values are univariate, one can rank the observed functions based on their depth values. 
These are defined as the \emph{sample depth-based ranks}, viz. for $i\in[n]$, we have that
$$\widehat{R}_i= \#\left\{X_{j}\colon \ID(h(X_{j }),h\circ\hat{\nu})\leq \ID(h(X_{i}),h\circ\hat{\nu}),\ j\in[n]\right\}\  .$$
It is shown in \cite{Ramsay2023b} that when $p=1$, the median of $\ID(X_{i}, \hat{\nu})$ is influenced by the covariance kernel of an observation $X_{i}$, which, along with the fact that conceptually, a function's depth is meant to differentiate between functions of different magnitude and shape \citep{Sun2011, DAI2020}, is the motivation for the use of depth in our methodology.  

The motivation for incorporating the derivative information stems from \cite{Hubert2012}, who showed that one can improve inference procedures for functional data by incorporating the derivatives of the observed functions into the depth values. 
This was employed also by \citet{Ramsay2023b}, which improved the performance of their hypothesis testing procedure. 
After computing the sample depth-based ranks $\widehat R_1,\ldots \widehat R_n$, the next phase of our procedure is to look for changes in the mean of these ranks.

In order to detect changes in these ranks, we employ the 
classical Kruskal--Wallis test statistic. 
Consider a candidate set of changepoints $\mathbf{r}=\{r_1,\ldots,r_m\}$, which we will always assume to be ordered by their indices, i.e., $r_0=0<r_1<\ldots<r_m<r_{m+1}=n$. 
Then the Kruskal--Wallis test statistic, where the ``treatment groups'' are the candidate segments $\{\widehat{R}_1,\ldots,\widehat{R}_{r_1}\}$, $\{\widehat{R}_{r_1+1},\ldots,\widehat{R}_{r_2}\}$, etc., is: 
\begin{equation}
    \mathcal{W}(\mathbf{r})=\frac{12}{n(n+1)} \sum_{j=1}^{m+1} (r_{j}-r_{j-1}) \overline{\widehat{R}}_{j}^{2}-3(n+1), \qquad\text{where}\qquad \overline{\widehat{R}}_{j}=\frac{1}{r_j-r_{j-1}}\sum_{i=r_{j-1}+1}^{r_j}\widehat{R}_i .
    \label{eqn:KW}
\end{equation}
Recall that in the Kruskal--Wallis ANOVA procedure, a large test statistic signals that there are differences between the groups. 
It follows that, in this setting, the bigger the differences in variability between segments, the bigger \eqref{eqn:KW} will be. 
Therefore, maximizing a version of \eqref{eqn:KW} over candidate changepoint sets $\mathbf{r}$ should give a set of time intervals in which the functions differ in variability. 

If the number of true changepoints $\ell$ is known, we can directly maximize $\cW$, which results in the estimate 
\begin{equation}\label{eqn::ellknown}
    \hat\bfk_\ell=\argmax_{1\leq r_1<\ldots<r_\ell<n}\cW(\bfr).
\end{equation}
When $\ell$ is known, it is often of interest to perform a hypothesis test for significance of the changepoints. 
In that case, we propose $\cW(\hat\bfk_\ell)=\sup_{1\leq r_1<\ldots<r_\ell<n}\cW(\bfr)$ as the test statistic, whose asymptotic distribution under the null hypothesis is a simple transformation of a standard Brownian bridge, see Theorem~\ref{thm::null} below. 

If the number of true changepoints $\ell$ is unknown, then maximizing the objective function $\cW$ over all possible candidate sets of changepoints is a degenerate problem. 
Therefore, we must add a penalty term, $\kappa_n>0$, on the number of changepoints, that is,
\begin{equation}    \label{eqn::pelt_kw}
\hat\bfk=\argmax_{\substack{ 0<r_1<\ldots<r_\ell<n}}[\cW(\bfr)-\ell\kappa_n].
\end{equation}
We call the estimates $\hat\bfk$ and $\hat\bfk_\ell$ the functional Kruskal--Wallis for covariance (FKWC) changepoint estimators. 
The estimate $\hat\bfk$ can be computed with the PELT algorithm \citep{Killick2012}, which, given the sample depth-based ranks, allows the changepoint estimates to be computed in linear time. 
The running time of our method then depends on our choice of $g$ and $\Dd$ in \eqref{eqn::integrated_depth}. 
For instance, if we let $N$ represent the number of points in the grid on which the observed functions are discretized when computing the depth values, and we take $\Dd$ to be the integrated dual depth or integrated-rank-weighted depth \citep{Cuevas2007,Ramsay2019}, then computing the set of random projection depth-based ranks takes $O(MN^2+M^2npd+n\log n)$ time.  
On the other hand, setting $\Dd$ to be the halfspace or simplicial depth exponential time in $d\times p$. 
Practically, it seems as though the majority of the computational burden comes in the form of computing the sample depth values. 
For example, 1 million univariate observations can be ranked in \texttt{R} (Version 4.4.0) in 0.37 seconds.\footnote{using an Intel(R) Core(TM) i7-8700K CPU @ 3.70GHz microchip and 32 Gb of RAM.} 
By contrast, computing the $\RP$ depth values of one million observations would take considerably longer with existing implementations in \texttt{R} (Version 4.4.0).


\section{Theoretical results}\label{sec::theory}
In this section, we present the theoretical results for the FKWC changepoint procedure. 
We first consider the setting where the number of changepoints is unknown, from which the setting where the number of changepoints is known is a simple consequence. 
\subsection{The number of changepoints is unknown}
Before presenting our main result, we first introduce a regularity condition on the multivariate depth function $\Dd$ in the integrated depth given by \eqref{eqn::integrated_depth}. 
We must have that $\Dd$ is \emph{$(K,\cF)$-regular}.
Let $\sB(\re^m,[0,1])$ denote the space of Borel functions $\re^m\to[0,1]$. 
For a family of functions, $\sF\subseteq\sB(\re^m,[0,1])$, define a pseudometric on $\cM_1(\re^m)$,  $d_\sF(\mu,\nu) = \sup_{f\in\sF}|\int_{\re^m} fd(\mu-\nu)|$, where $\mu,\nu\in\cM_1(\re^m)$. 
\begin{definition}[\cite{ramsay2022concentration}]\label{def::kf-reg}
We say that $\Dd$ is \emph{$(K,\sF)$-regular} if there exists some class of functions $\sF\subset \sB(\re^m,[0,1])$ and a positive constant $K>0$ such that $\Dd(x,\cdot)$ is $K$-Lipschitz with respect to the $\sF$-pseudometric uniformly in $x$, i.e., for all $\mu,\nu\in\cM_1(\re^m)$
\[
\sup_{x\in\re^m}|\Dd(x,\mu)-\Dd(x,\nu) | \leq K d_\cF(\mu,\nu).
\]
\end{definition}
\noindent Given let class of functions $\sF$, $\VC(\sF)$ be the Vapnik--Chervonenkis dimension of $\sF$ \citep{Vapnik1999-ee}. 
We now introduce the following condition on the integrated depth given in \eqref{eqn::integrated_depth}.
\begin{condition}\label{cond::depth}
There is some $K>0$ such that:
\begin{enumerate}
    \item For all $m\geq 1$, there exists some family of functions $\sF\subseteq \sB(\re^m,[0,1])$ with $\VC(\sF)<\infty$ such that $\Dd$ is $(K,\sF)$-regular.
    \item For all $i\in [n]$, the function $f(x)=\Prr{\ID(X_i,\nu)\leq x}$ is $K$-Lipschitz.
\end{enumerate}
\end{condition}
\noindent Many common multivariate depth functions, including halfspace depth, simplicial depth, spatial depth and integrated-rank-weighted depth are $(K,\cF)$-regular. 
For such depths, $K$ is often constant and at most linear in $m$ and $\VC(\cF)$ is at least linear in $m$, and at most $O(m^2\log m)$ \citep{ramsay2022concentration}. 
For instance, halfspace depth is $(1,\cF)$-regular for $\cF$ such that $\VC(\cF)=O(m)$. 
Item 2.\ in Condition~\ref{cond::depth} is a smoothness condition on the cumulative distribution function of the depth values (computed with respect to the population depth).

Let $\tau_i=\Prr{\ID(h( X_{k_i}),h\circ\nu)\leq \ID(h( X_{k_i+1}),h\circ\nu)}-1/2$ be the size of the $ith$ changepoint (see Remark \ref{rem::change_capt}) and let $\tau_n =\min_{i\in[\ell]}\tau_i$. 
Next, let $\Delta_n=\min_{i\in[\ell+1]} k_i-k_{i-1}$ be the minimum spacing between changepoints, $\hat\ell=|\hat\bfk|$ and let $\lambda_{n,\cF,K}= K^3\ell\sqrt{\VC(\cF)}\log n$. 
The following gives the rate of the convergence of the proposed changepoint estimates. 
\begin{theorem}\label{thm::main-result}
    If Conditions \ref{cond::contin} and \ref{cond::depth} hold, then there exists universal constants $c_1,c_2,c_3>0$ such that if $\kappa_n\leq c_1\Delta_n \tau_{n}^2/\ell$ and $\lambda_{n,\cF,K}=o(\kappa_n)$, then there exists $n_0$ such that for all $n\geq n_0$ and $d,p\geq 1$, it holds that
    $$\Prr{\left\{\hat\ell=\ell\right\} \cap \left\{\max_{\hat k_j\in\hat\bfk}\min_{k_i\in\bfk}|k_i-\hat k_j|\leq c_2\frac{\lambda_{n,\cF,K}}{\tau_n^2}\right\}}\geq 1-c_3/n.$$
\end{theorem}
The proof of Theorem~\ref{thm::main-result} is deferred to Appendix \ref{sec::Proofs}. 
For $a,b\in\re$, we say $a\lesssim b$ ($a\gtrsim b$) whenever there exists a universal constant $C>0$ such that $a\leq C b$ ($a\geq C b$). 
Note that we do not assume that the number of changepoints, the distance between the changepoints, or the jump sizes are fixed in $n$. We only require that $\Delta_n \tau_{n}^2\gtrsim \ell^2\lambda_{n,\cF,K}$, which is similar to requirements in the univariate setting, \citep{Wang2020, Padilla2021}.
In fact, in the univariate setting, no algorithm can detect a general change in distribution if $\tau_n(\Delta_n)^{1/2}\lesssim 1$ \citep{Padilla2021}. 
Therefore, (for $\ell\lesssim\log n$), the implied bound on the signal-to-noise ratio is necessary, up to the logarithmic terms.
In the setting of $p,d<<n$\footnote{Here, $a<<b$ ($a>>b$) is taken to be $a$ must less (more) than $b$.}, Theorem~\ref{thm::main-result} yields that $\hat\bfk$ achieves the minimax localization rate for detecting general changes in distribution in the univariate setting \citep{Padilla2021}, up to logarithmic terms, when $\Delta_n \tau_{n}^2\gtrsim \ell^2\lambda_{n,\cF,K}$. 
Therefore, this procedure is optimal (up to logarithmic terms) for detecting changes in the medians of the depth values computed with respect to the population depth function, i.e.,  $\ID(h( X_{k_i}),h\circ\nu)$, for fixed $p$ and $d$.

Recall that among the popular choices for $\Dd$, the smallest values of $K$ and $\VC(\cF)$ are $1$ and $O(m)$, respectively. 
In this case, observe that the localization rate becomes $O(\ell\sqrt{pd}\log n/\tau^2_n)$, and so the procedure will work better for low to moderate levels of $d$ and $p$, such as in the case of f-MRI data. 

Note that Theorem~\ref{thm::main-result} has a minor dependence on the parameter $d$ because we have assumed that the functions are observed fully. 
If the functions are instead observed discretely and sparsely over their domain, then we expect the convergence rate given in Theorem~\ref{thm::main-result} to be optimistic, as one would first have to smooth the functions, which becomes increasing difficult as $d$ increases, see Remark \ref{rem::full_curves}. 
However, in this work, we are concerned with the case where $d$ is small $(\leq 4)$ and the functions are observed densely. 

\begin{remark}\label{rem::change_capt}
The type of change captured by the procedure is entirely encapsulated in the following condition: 
If there exists a changepoint at time $k_i$, then $\Prr{\ID(h( X_{k_i}),h\circ\nu)\leq \ID(h( X_{k_i+1}),h\circ\nu)}\neq 1/2$.
Therefore, we must have that the change in the distribution at $k_i+1$ must imply a location change in the depth value sequence. 
Definitionally, depth functions distinguish between functions of differing magnitudes and/or shapes \citep{Sun2011, DAI2020}, so we say this is equivalent to a change in variability.  
In addition, if $\ID$ is taken to be the random projection depth, a fairly large class of changes in the covariance kernel imply changes in the mean of the sequence $\ID(X_i,\nu)$ for a large class of distributions $\nu$ \citep{Ramsay2023b}. 
We expect such changes to also change the sequence $\ID(h(X_i),h\circ\nu)$, which is confirmed by our simulation study (see Section \ref{sec::sim}) and the simulation study of \citet{Ramsay2023b}. 
\end{remark}
\begin{remark}\label{rem::full_curves}
Condition~\ref{cond::contin} implies that we have access to the observed functions $X_i$ in full. In reality, we often only have access to $X_i$, observed noisily at a finite number of points, say $m_i$. 
These discretized functions are then smoothed prior to analysis. 
Convergence rates of the integrated depths (and other depths) computed on noisy discretized functions were characterized by \citet{Nagy2019}. 
Essentially, under some broad, technical assumptions on the curves, in the setting of $m_i>>n$, which we can describe as ``dense'' functional data, where  the uniform convergence rate of an integrated depth function to the population depth function is not degraded \citep[][see Theorem 6 and Section 5]{Nagy2019}. 
Therefore, if the curves are densely observed and are sufficiently smooth, we need not be concerned about this point. 
On the other hand, for sparse functional data, we expect the rate of convergence of the FKWC changepoint estimates to be slower than that as given in Theorem~\ref{thm::main-result}. 
\end{remark}
\begin{remark}\label{rem::mv_est}
The algorithm described is similar to that of Algorithm 2 of \cite{Ramsay2023b} for multivariate data. It is natural to ask if the theoretical analysis of \cite{Ramsay2023b} applies here. 
However, the rate given in Theorem 2 of \cite{Ramsay2023b} is slower than that given in Theorem~\ref{thm::main-result}. 
Therefore, applying their analysis would produce a weaker result. 
Furthermore, Theorem~\ref{thm::main-result} offers an improvement over Theorem 2 of \cite{Ramsay2023b}. 
It is straightforward to show that the result of Theorem~\ref{thm::main-result} extends to the multivariate case, which means Theorem~\ref{thm::main-result} implies a faster rate of convergence for the multivariate changepoint estimator of \cite{Ramsay2023b}. 
\end{remark}

\subsection{The number of changepoints is known}
We now consider the setting where the number of changepoints is known. 
The fact that the hypothesis tests and changepoint estimates $\hat\bfk_\ell$ are consistent follows from the analysis of Theorem \ref{thm::main-result}. 
Let $B(t)$ be a standard Brownian bridge. 
\begin{corollary}\label{cor::ell_known}
 If Conditions \ref{cond::contin}--\ref{cond::depth} hold, then there exists universal constants $c_1,c_2>0$ and there exists an integer $n_0$ such that for all $n\geq n_0$ and $d,p\geq 1$, it holds that
    $$\Prr{\max_{\hat k_j\in\hat\bfk_\ell}\min_{k_i\in\bfk}|k_i-\hat k_j|\leq c_1\frac{\lambda_{n,\cF,K}}{\tau_n^2}}\geq 1-c_2/n.$$
    Furthermore, if $\ell>0$, then for any universal constant $C>0$, we have that $\Prr{\cW(\hat\bfk_\ell)\geq C}\to 1$ as $n\to\infty$.
\end{corollary}
\noindent The proof follows directly from the last paragraph of the proof of Theorem~\ref{thm::main-result}. 
We also present the behavior of our proposed test statistics under the null hypothesis for the AMOC and the epidemic setting. 
\begin{theorem}\label{thm::null}
Suppose that Conditions \ref{cond::contin}--\ref{cond::depth} hold and that $\ell=0$.
Then it holds that 
$$\sup_{1\leq r<n}\cW(r)\cond \sup_{0<t<1} t(1-t)|B(t)|^2,$$
and that
\begin{equation*}
\sup_{0<r_1<r_2<n}\cW(r_1,r_2)\cond \sup_{t_1,t_2\in (0,1)}\left(\frac{1}{(t_{2}-t_{1})(1-t_{2}+t_{1})}\right)(B(t_2)-B(t_1))^2.
\end{equation*}
\end{theorem}
\noindent The proof of Theorem~\ref{thm::null} is deferred to Appendix~\ref{sec::Proofs}. 
\section{Simulation}\label{sec::sim}
\begin{figure}[t!]
\begin{minipage}[c]{.32\textwidth} 
    \centering
    \includegraphics[width=1\textwidth]{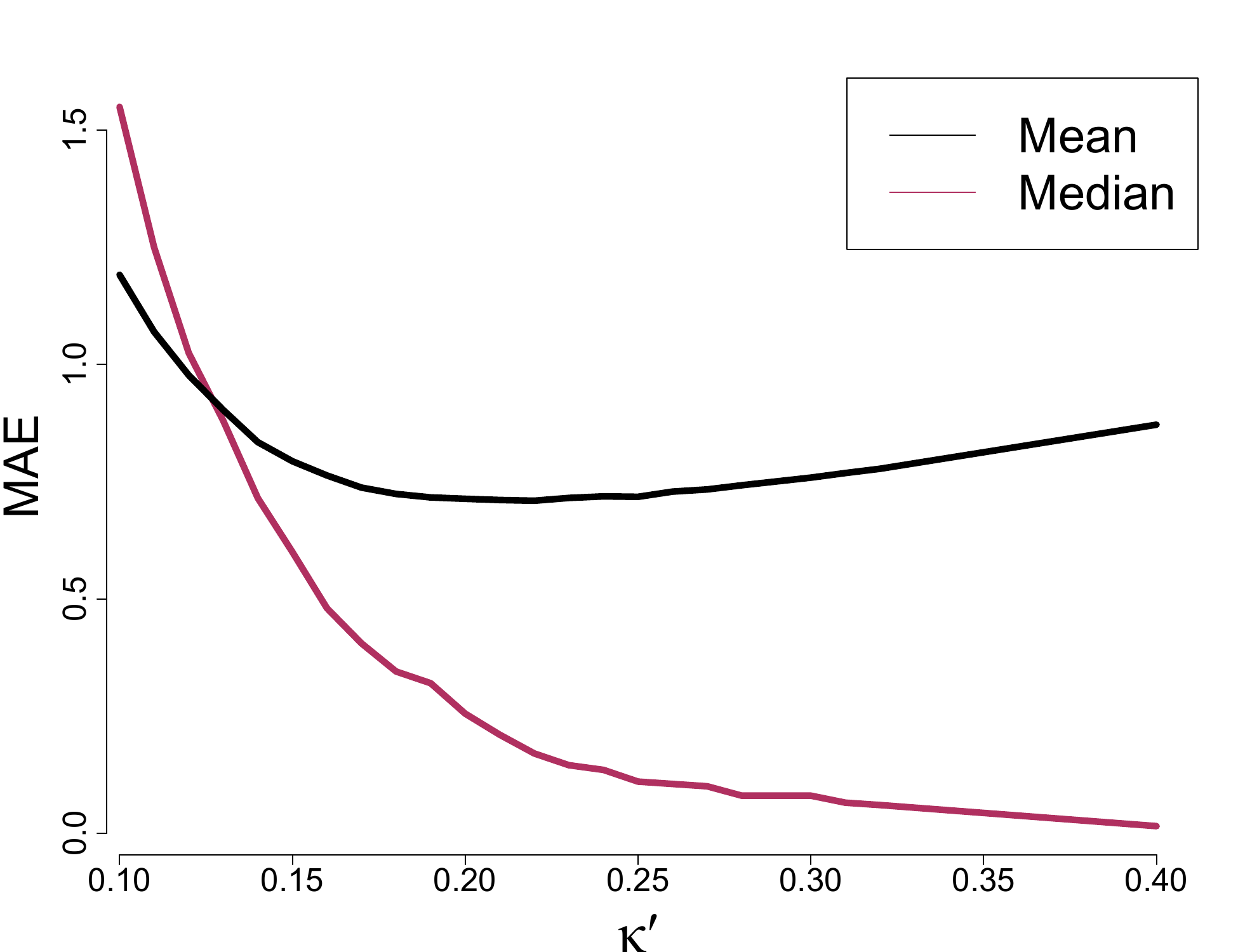}
     \caption*{(a) }
\end{minipage}
\begin{minipage}[c]{.32\textwidth} 
    \centering
  \includegraphics[width=1\textwidth]{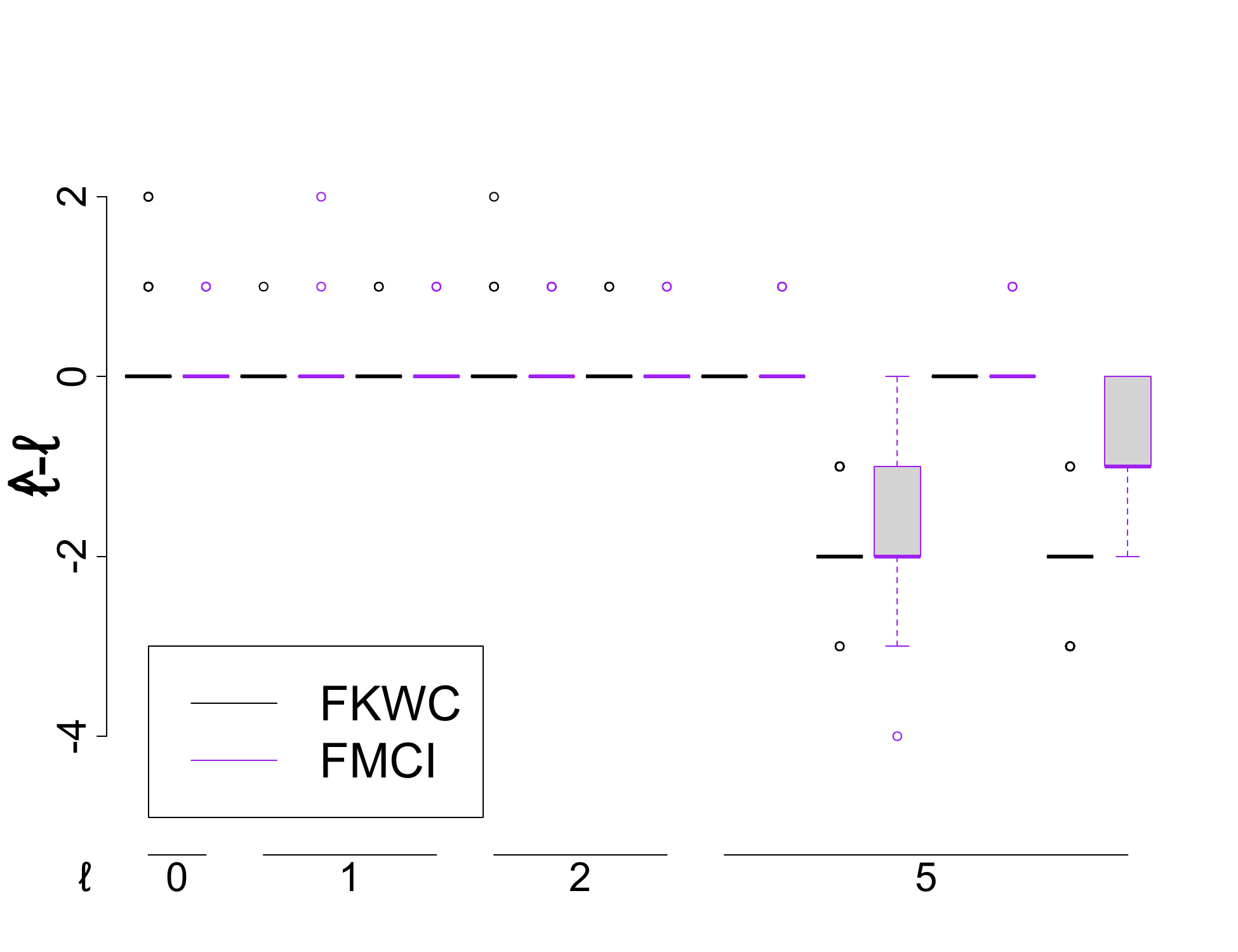}
    \caption*{(b)  }
\end{minipage}
\begin{minipage}[c]{.32\textwidth} 
    \centering
  \includegraphics[width=1\textwidth]{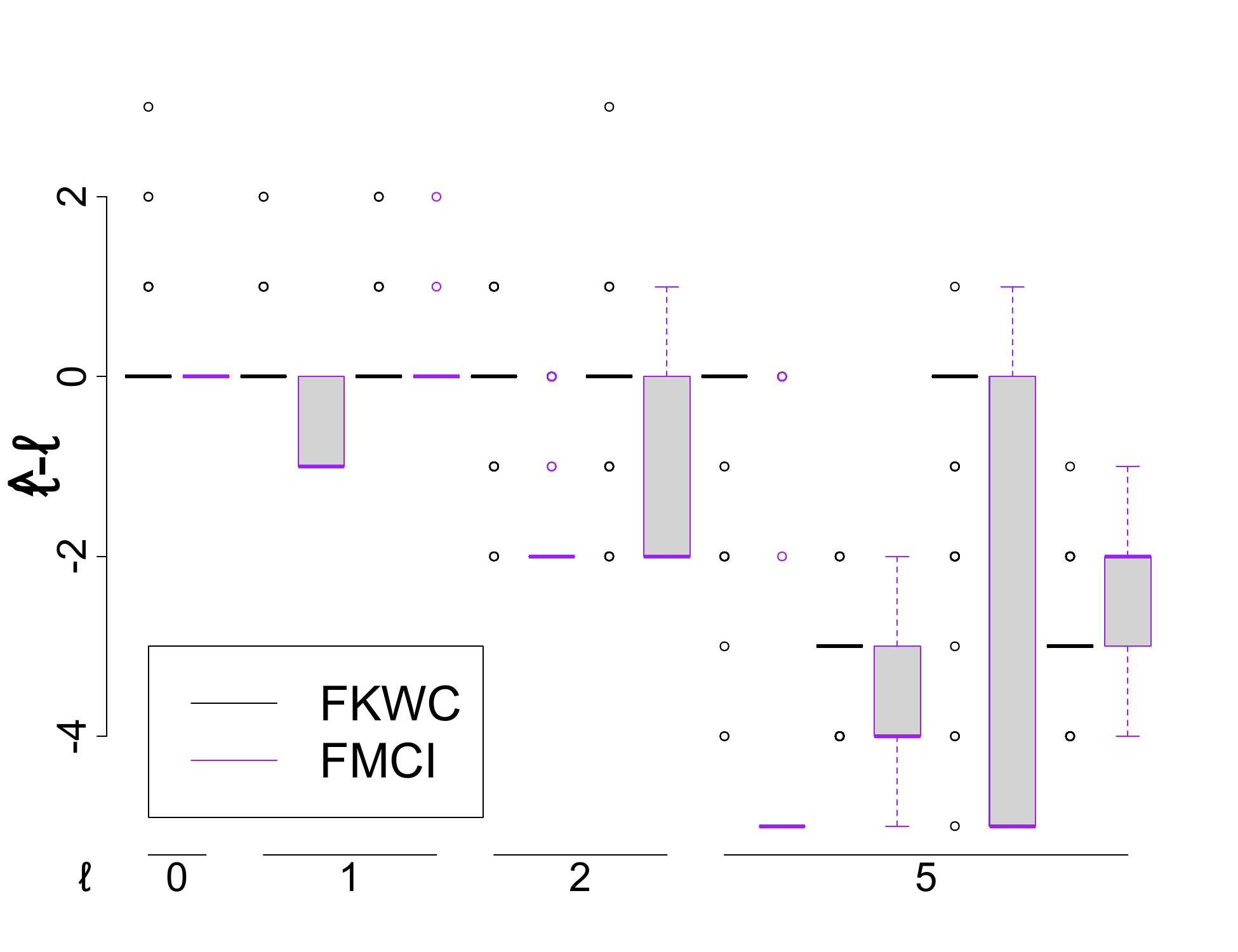}
    \caption*{(c)  }
\end{minipage}
 \caption{(a) Mean and median of the mean absolute error $|\ell-\hat{\ell}|$ over all simulation scenarios for different values of $\kappa'$ for $n=500$. We can see that choosing $\kappa'\in (0.2,0.3)$ produces low levels of error. (b) Boxplots of $\hat{\ell}-\ell$ when the sequence of data contained different amounts of changepoints (as labelled on the horizontal axis) under the Gaussian simulation scenario. (c) Boxplots of $\hat{\ell}-\ell$ when the sequence of data contained different amounts of changepoints under the Student $t$ simulation scenario. We can see that the performance of the methods are similar when the data is Gaussian, and the FKWC method performs better when the data is Student $t$. }
  \label{fig::Param_choice}
\end{figure}
\subsection{Simulation setup}
In order to test the empirical performance of our methodology and compare to existing methods, we simulated observations from several changepoint models. 
For brevity, we restricted the simulation to the case where $d=p=1$. 
In each simulation run, $\nu^1,\ldots,\nu^{\ell+1}$ were such that all observed functions were either Gaussian processes, denoted $\cG$, Student $t$ processes with degrees of freedom equal to three, denoted $t_3$, or Skewed Gaussian processes denoted  $\mathcal{SG}$. 
Each $\nu^i$ had a squared exponential covariance kernel 
$$\kerr_{i}(s,t; \alpha_i,\beta_i)=\beta_i\ e^{\frac{-(s-t)^2}{2\alpha_i^2}}.$$
At each changepoint, either $\alpha_i\neq \alpha_{i+1}$ (shape change) or $\beta_i\neq \beta_{i+1}$ (scale change). 
Changes in $\alpha$ correspond to a `shape' change in the data, while changes in $\beta$ correspond to a scale, i.e. a magnitude, change in the data. 
We considered the following scenarios:
\begin{itemize}
    \item AMOC: We simulated data with zero changepoints as well as with one changepoint in the middle of the sample. Sample sizes of 100, 200, and 500 were used.
    \item Epidemic: We simulated data with two uniformly random changepoints, where we required that the changepoints were at least 10\% of the sample size apart. Sample sizes of 100, 200, and 500 were used.
    \item Multiple: We simulated data with five randomly placed changepoints which also had to be at least 10\% of the sample size apart. Sample sizes of of 200, 500, 1000 and 2500 were used. We ran four different simulation scenarios, two where the changepoints were shape-type and two where the changes were magnitude-type. 
Within these groups, the set of changepoints was either ``ascending'', i.e., $\alpha$ or $\beta$ was increasing with each change or ``alternating'', i.e., $\alpha$ or $\beta$ was oscillating between a high and low value with each change. 
\end{itemize}
We ran all of the above cases once where the changepoints were shape changes and once where they were scale changes. 

In terms of comparing the FKWC procedure under the two depths $\RP$ and $\MFHD$, the performance of each was very similar, with $\MFHD$ being slightly better for magnitude differences and $\RP$ being slightly better for shape differences. 
Therefore, we only present the results with $\ID=\RP$.  


\subsection{The case where $\ell$ is unknown}
The purpose of the simulation study was to find a suitable choice for the tuning parameter $\kappa_n$, as well as to compare the performance of the FKWC changepoint method to existing methods. 
We first discuss choosing the value of $\kappa_n$. 
It was observed by \cite{RAMSAY2023} that $\kappa_n\in(3.74+0.15 \sqrt{n},3.74+0.25\sqrt{n})$ performs well in the multivariate setting. 
In this study, we tested $\kappa_n=3.74+\kappa' \sqrt{n}$ for $\kappa'\in(0.1,0.4)$ to see if the same parameter settings apply to the functional data setting. 
We ran the PELT algorithm on the simulated data for all of the scenarios, i.e., for data which had 0, 1, 2 and 5 changepoints. 
Figure \ref{fig::Param_choice} shows the average and median of the mean absolute errors in the estimated amount of changepoints, i.e., $|\ell-\hat{\ell}|$, over all simulation runs for different values of $\kappa'$ when $n=500$. 
Figure \ref{fig::Param_choice} shows that the best values of $\kappa'$ are in the range $0.2-0.3$. 
Note that these values are higher than that of the multivariate setting \citep{RAMSAY2023} and that the algorithm becomes less sensitive to the choice of $\kappa'$ as the sample size increases. 


\begin{figure}[t!]
\begin{minipage}[c]{.49\textwidth} 
    \centering
    \includegraphics[width=0.9\textwidth]{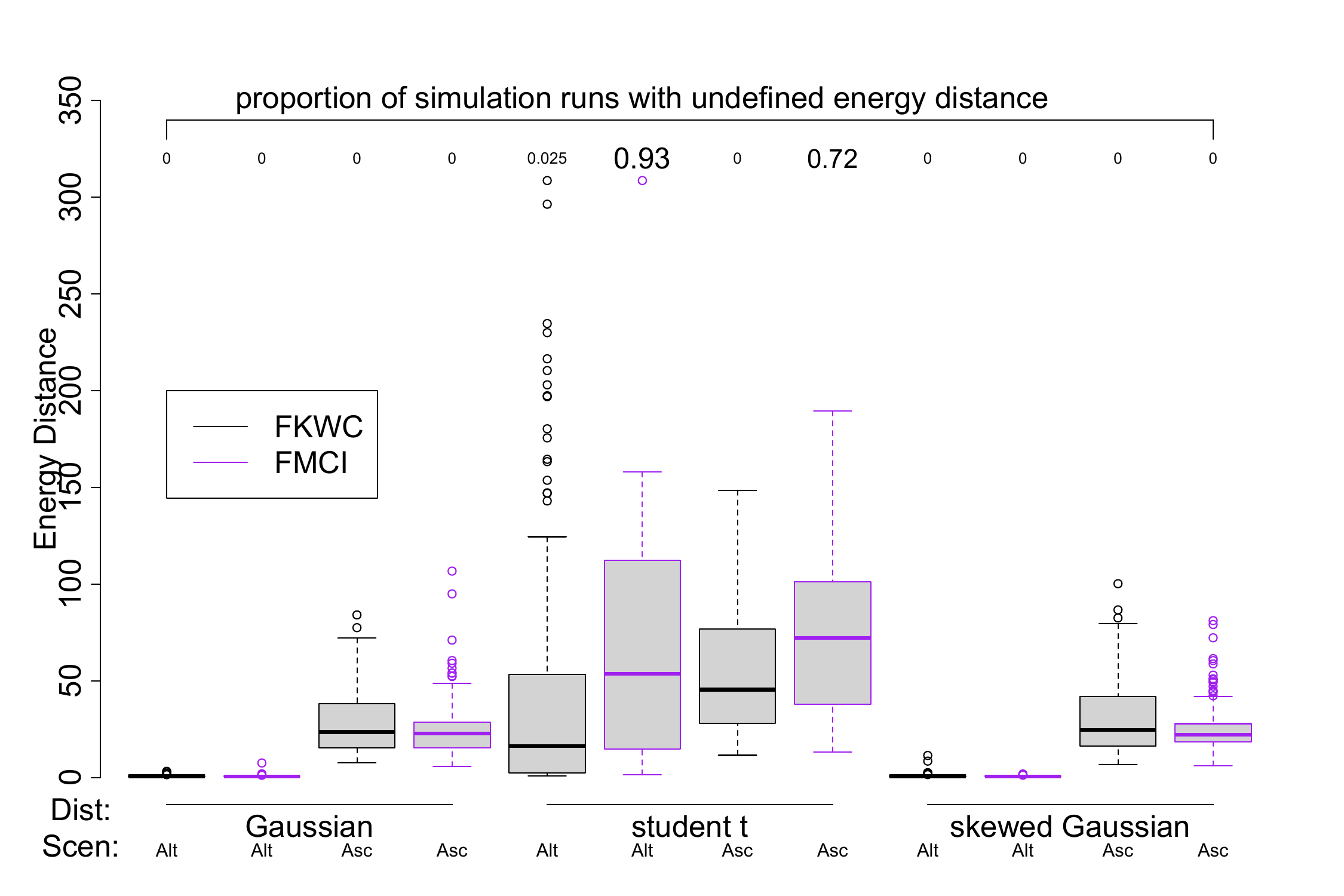}
     \caption*{Scale changes}
\end{minipage}
\begin{minipage}[c]{.49\textwidth} 
    \centering
    \includegraphics[width=0.9\textwidth]{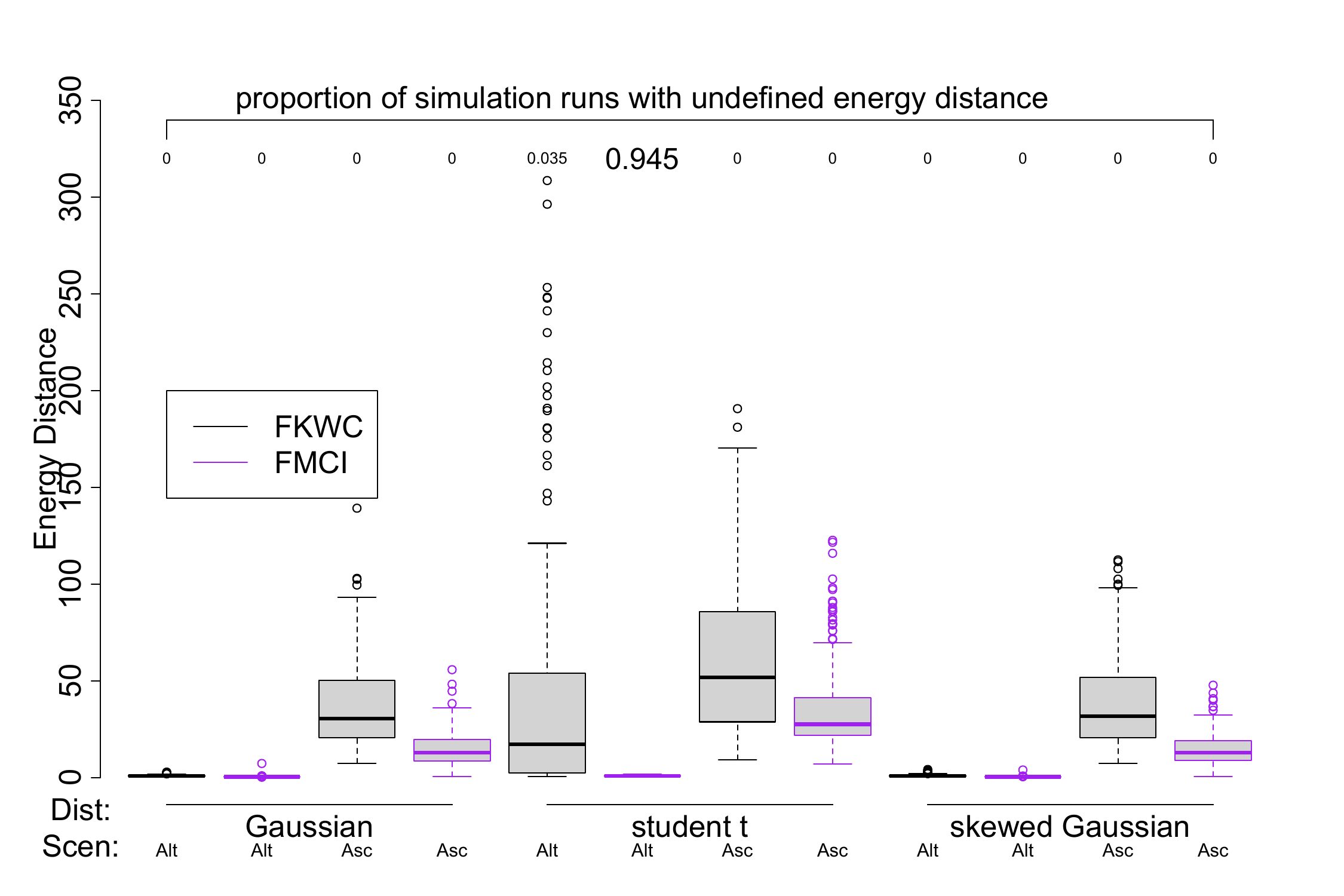}
    \caption*{Shape changes}
\end{minipage}
 \caption{Energy distance between the estimated and true changepoints when $n=500$ for the FKWC procedure and the FMCI method. The FKWC method was used with $\kappa'=0.3$. Only the runs in which there were five changepoints are represented. The numbers at the top of the graph are the proportion of runs in which the algorithm failed to identify any changepoints. The labels on the horizontal axis represent the distribution of the underlying data and the changepoint scenario; either alternating or ascending. }
  \label{fig::Energy_Dist}
\end{figure}


We now compare the FKWC changepoint estimator (with $\ID=\RP$) to the FMCI method of \citet{Harris2021}. We chose to compare to \citet{Harris2021} because of FMCI's ability to detect multiple changepoints, as well as its accessible implementation and computational speed. 
We use the default parameters kindly provided in the \texttt{fmci} package. 
Note that the FMCI method can detect changes in both the mean function and covariance kernel, whereas the FKWC procedure can only detect changes in variability.
It should be noted that we used the same simulation to evaluate the best parameter choice for the FKWC method, and so the results could be biased in favor of the FKWC procedure. 
We do not feel this plays a major role in the conclusions drawn from the comparison. 

Figure \ref{fig::Param_choice} contains boxplots of $\hat{\ell}-\ell$ for different data models in the simulation. 
The results of skewed Gaussian and Gaussian were similar so we only present those of Gaussian and Student $t$. 
We see that the results of both methods are similar under the Gaussian setting (while slightly favoring the FMCI method), but favor the FKWC procedure when the data is heavy-tailed. 
Both methods tend to underestimate the number of changepoints in the heavy-tailed case.  
We observed that both of these methods had more difficulty in the ascending scenarios, i.e., the simulation runs where either the $\alpha$ or $\beta$ parameters were increasing at each changepoint. 

To evaluate the accuracy of the algorithms, we also look at the energy distance between  the estimated changepoint set and true changepoint set for each method. 
The energy distance between the estimated and the true changepoint set can be written as 
$$\frac{2}{\hat{\ell}\ell}\sum_{i=1}^{\hat{\ell}}\sum_{j=1}^{\ell}|\hat{k}_i-k_j|-\frac{1}{\hat{\ell}^2}\sum_{i=1}^{\hat{\ell}}\sum_{j=1}^{\hat{\ell}}|\hat{k}_i-\hat{k}_j|-\frac{1}{\ell^2}\sum_{i=1}^{\ell}\sum_{j=1}^{\ell}|k_i-k_j|.$$
We use this distance because, as discussed in Appendix B.3 of \citet{Harris2021}, the energy distance measures the average error in estimating each changepoint, rather than the error of the most poorly estimated changepoint in the set. 
One criticism is that if the algorithm fails to identify any changepoints, then the energy distance to a set of true changepoints will not be defined. 

Figure \ref{fig::Energy_Dist} shows boxplots of the energy distance between the estimated changepoint set and true changepoint set, for each method. 
The numbers along the top of the graph indicate the proportion of simulation runs in which the algorithm failed to identify any changepoints. 
Figure \ref{fig::Energy_Dist} shows that the FMCI method performs better in the Gaussian and skewed Gaussian scenarios when the changepoints are `ascending'. 
However, the FMCI method can perform poorly in the heavy-tailed scenario. 
For instance, for heavy-tailed processes with scale changes, the FMCI failed to detect a changepoint in the majority of the simulation runs. 
We conclude that the FMCI method performs better when the data are not heavy-tailed, but the FKWC performs better when the data are heavy-tailed. 
In other words, the FKWC procedure sacrifices some of the accuracy of the FMCI method for robustness. 




\subsection{The case where $\ell$ is known}

We now evaluate the performance of the FKWC hypothesis tests for AMOC and epidemic alternatives. 
We only present the results from the AMOC alternative, since the results from the epidemic alternatives resulted in the same conclusions. 
The results from the epidemic alternatives can be seen in Appendix \ref{app::sim}. 
All tests were carried out at the 5\% level of significance. 
For the AMOC alternative, we compare our methods to the methods of \citet{Sharipov2019, Dette2020a}. 
The code for these methods was kindly provided by the authors.  
We used 200 bootstrap samples with a block length of 1 for both of the competing methods. 
We only report the results for the integrated test of \citet{Sharipov2019}, since it had a higher power than the other test proposed in their paper. 
For the method of \citet{Dette2020a}, we used 49 b-spline basis functions to smooth the data first and, note that using a Fourier basis resulted in slightly lower power. 
We did not smooth the data for use with the method of \citet{Sharipov2019}.

\begin{figure}[t!]
\begin{minipage}[c]{.49\textwidth} 
    \centering
    \includegraphics[width=0.9\textwidth]{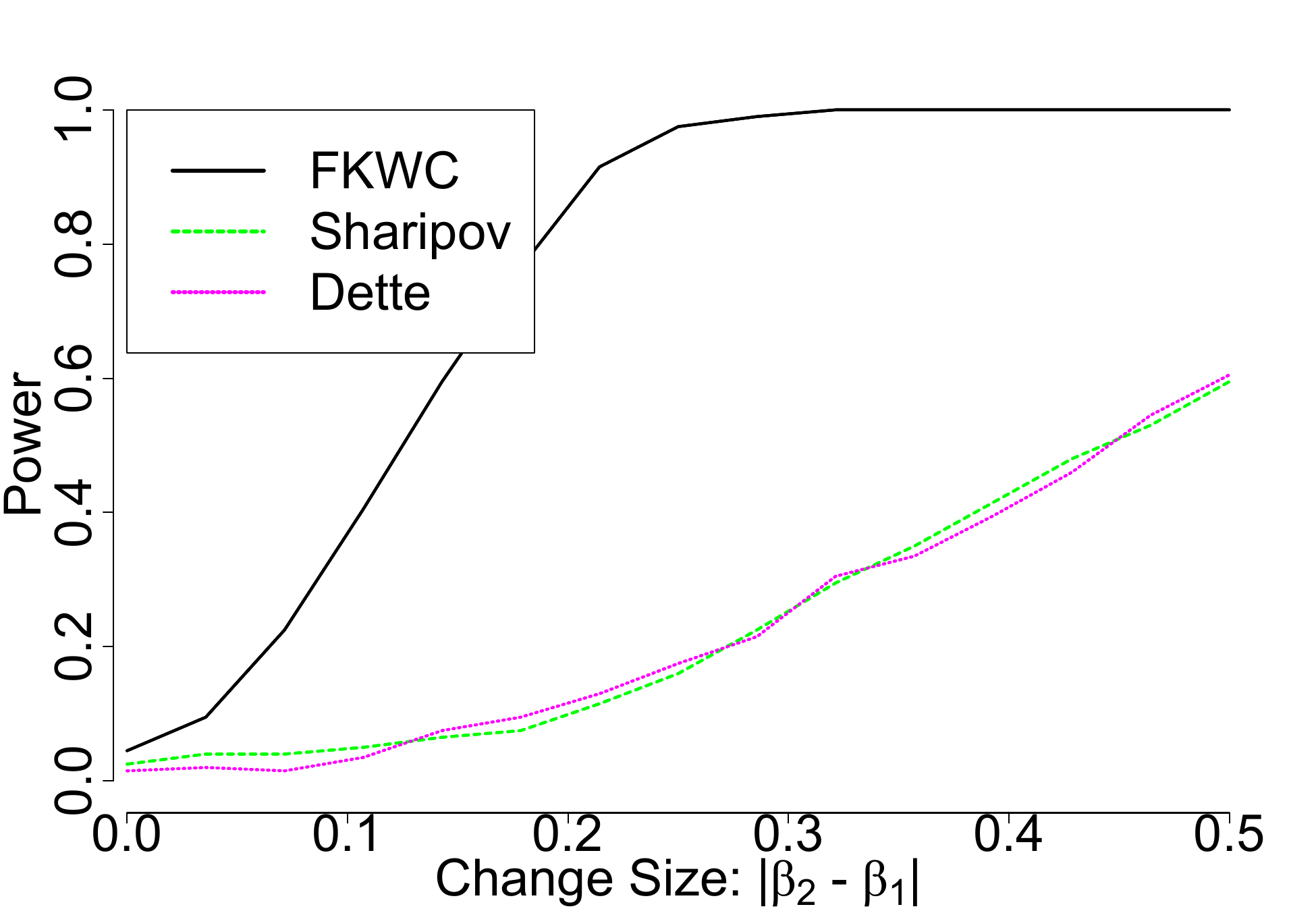}
     \caption*{(a) Gaussian processes, scale change }
\end{minipage}
\begin{minipage}[c]{.49\textwidth} 
    \centering
    \includegraphics[width=0.9\textwidth]{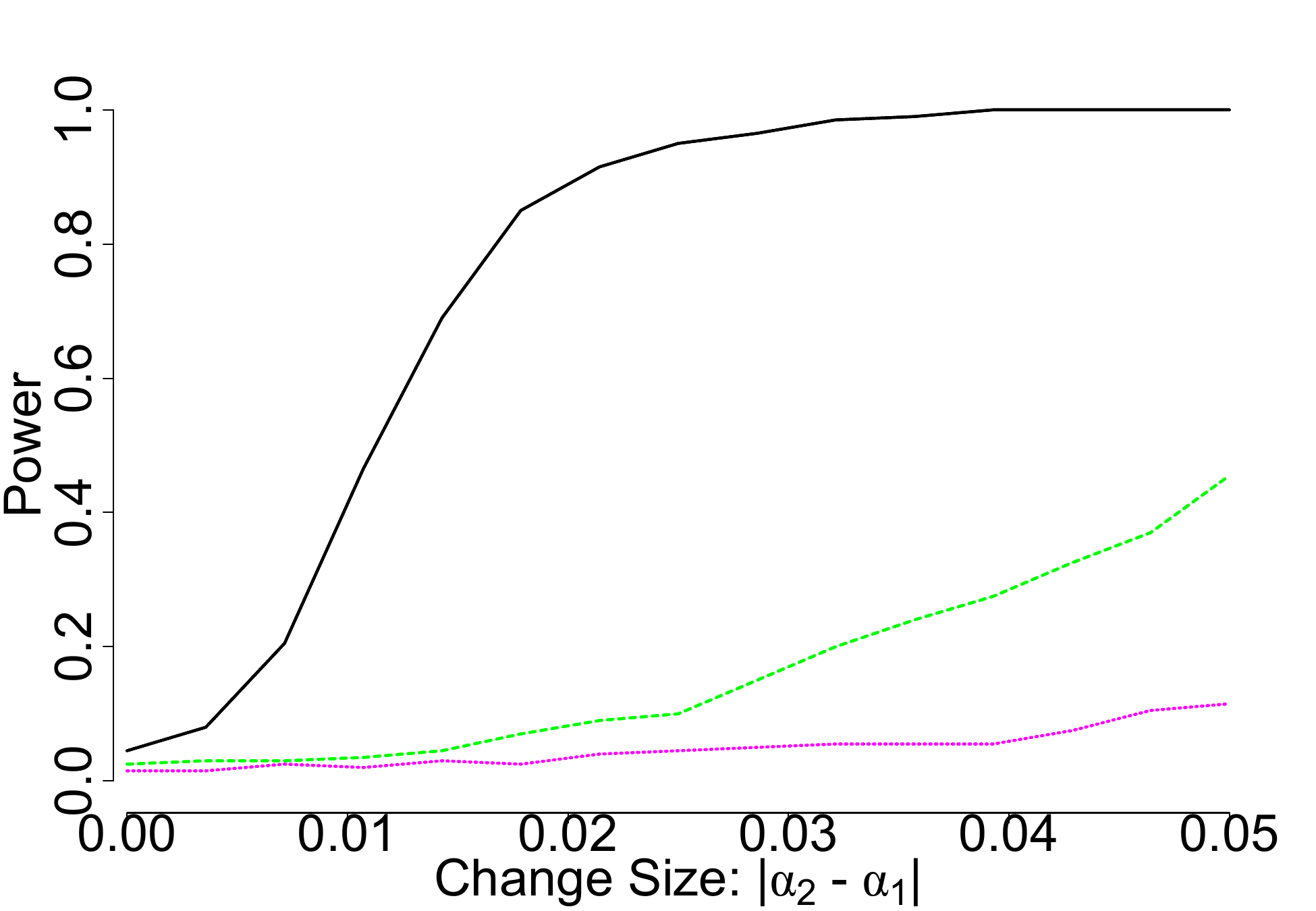}
    \caption*{(b) Gaussian  processes, shape change }
\end{minipage}
\hfill\newline
\begin{minipage}[c]{.49\textwidth} 
    \centering
    \includegraphics[width=0.9\textwidth]{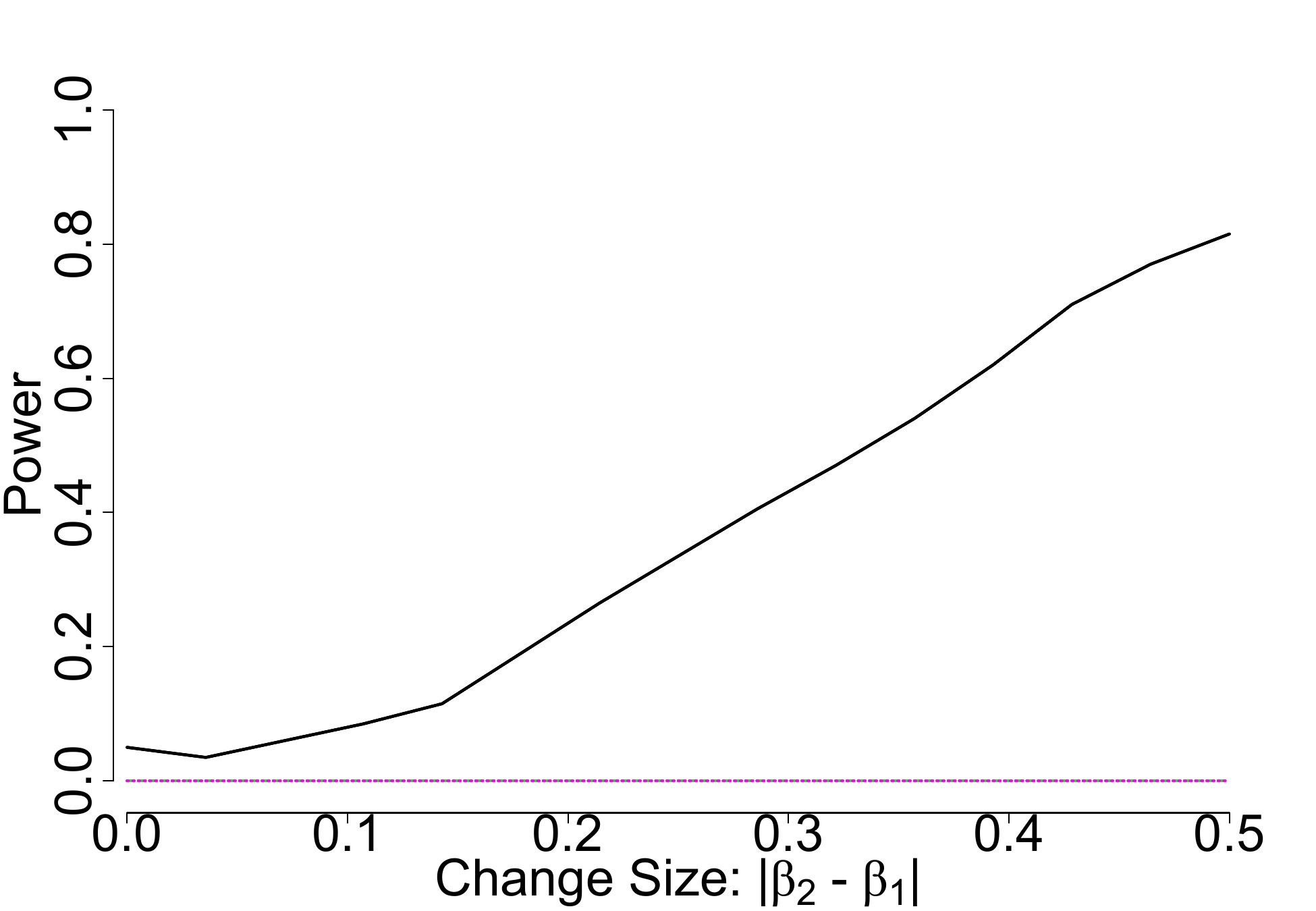}
    \caption*{(c) Student $t_3$  processes, scale change }
\end{minipage}
\begin{minipage}[c]{.49\textwidth} 
    \centering
    \includegraphics[width=0.9\textwidth]{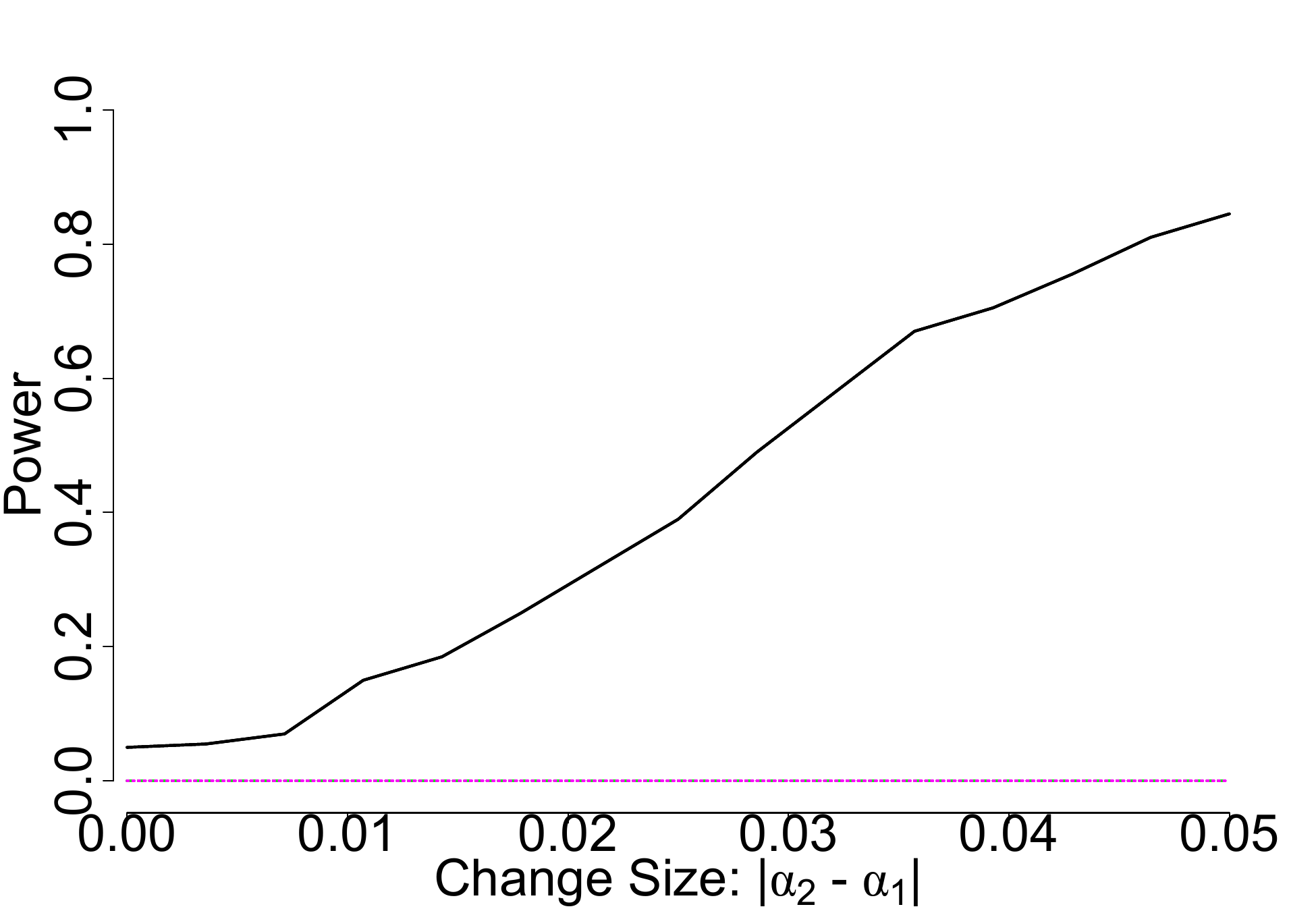}
    \caption*{(d)  Student $t_3$  processes, shape change}
\end{minipage}
  \caption{Empirical power curves of the FKWC procedure and the methods of \citet{Sharipov2019, Dette2020a}, under different data models. Here, we consider the AMOC tests and $n=100$. We see that the FKWC has higher power for both the Gaussian and Student $t$ scenarios.}
  \label{fig:AMOC_other}
\end{figure}

Table \ref{tab::AMOC_sizes} gives the empirical sizes of the hypothesis tests when $n=100$. 
Observe that for all tests, the empirical sizes were less than or equal to the nominal size of 0.05 for all methods.
Figure \ref{fig:AMOC_other} compares the empirical power of the methods of \citet{Sharipov2019, Dette2020a} to the empirical power of the FKWC procedure as the change size grows for $n=100$. 
It can be seen in Figure \ref{fig:AMOC_other} that the FKWC test has a higher power than its competitors for the data models in this simulation study. 
It is also apparent that the heavy-tailed processes completely corrupt the competing methods. 
Although the FKWC methods have higher power than competing methods under these data models, the competing methods have some features that the FKWC methods do not. 
These methods are theoretically sound for dependent data and the method of \citet{Dette2020a} can test for ``relevant'' changes in the covariance operator, rather than the standard hypothesis of any change in the covariance operator and/or variability.

We also tested the performance of the FKWC procedure when the data were dependent. 
We simulated functions from the autoregressive model as discussed in the simulation section of \cite{Sharipov2019} and ran the FKWC test on those time series. 
Table \ref{tab::shar_ar} in Appendix \ref{app::sim} shows the results under this model. 
We see that the FKWC procedure has higher power than competing methods, though they tend to have higher empirical sizes.
Here, the FKWC procedure with $\RP$ performed much better than the FKWC procedure with  $\MFHD$. 
They had similar empirical powers, but the $\RP$ version had an empirical size of 0.06, compared to 0.10 for the $\MFHD$ version. 
Overall, the performance of the FKWC test is better than its competitors under these simulation models. 

\begin{table}[t]
\centering
\caption{Empirical sizes of the different AMOC hypothesis tests when $n=100$, under Gaussian and heavy-tailed data. We see that all tests have empirical size less than or equal to the nominal size of $0.05$. }
\begin{tabular}{rrrrr}
  \hline
 Change type &\multicolumn{2}{c}{Scale} & \multicolumn{2}{c}{Shape}\\ \hline
 & Gaussian $\cG$ & Student $t_3$ & Gaussian $\cG$ & Student $t_3$\\ 
  \hline
FKWC & 0.04 & 0.05 & 0.04 & 0.05 \\ 
  Sharipov & 0.02 & 0.00 & 0.02 & 0.00 \\ 
  Dette & 0.01 & 0.00 & 0.01 & 0.00 \\ 
   \hline
\end{tabular}
\label{tab::AMOC_sizes}
\end{table}
\section{Data analysis}\label{sec::DA}
\subsection{Changes in volatility of social media intraday returns}
\begin{figure}[t]
\begin{minipage}[c]{.5\textwidth}
    \centering
    \includegraphics[width=\textwidth]{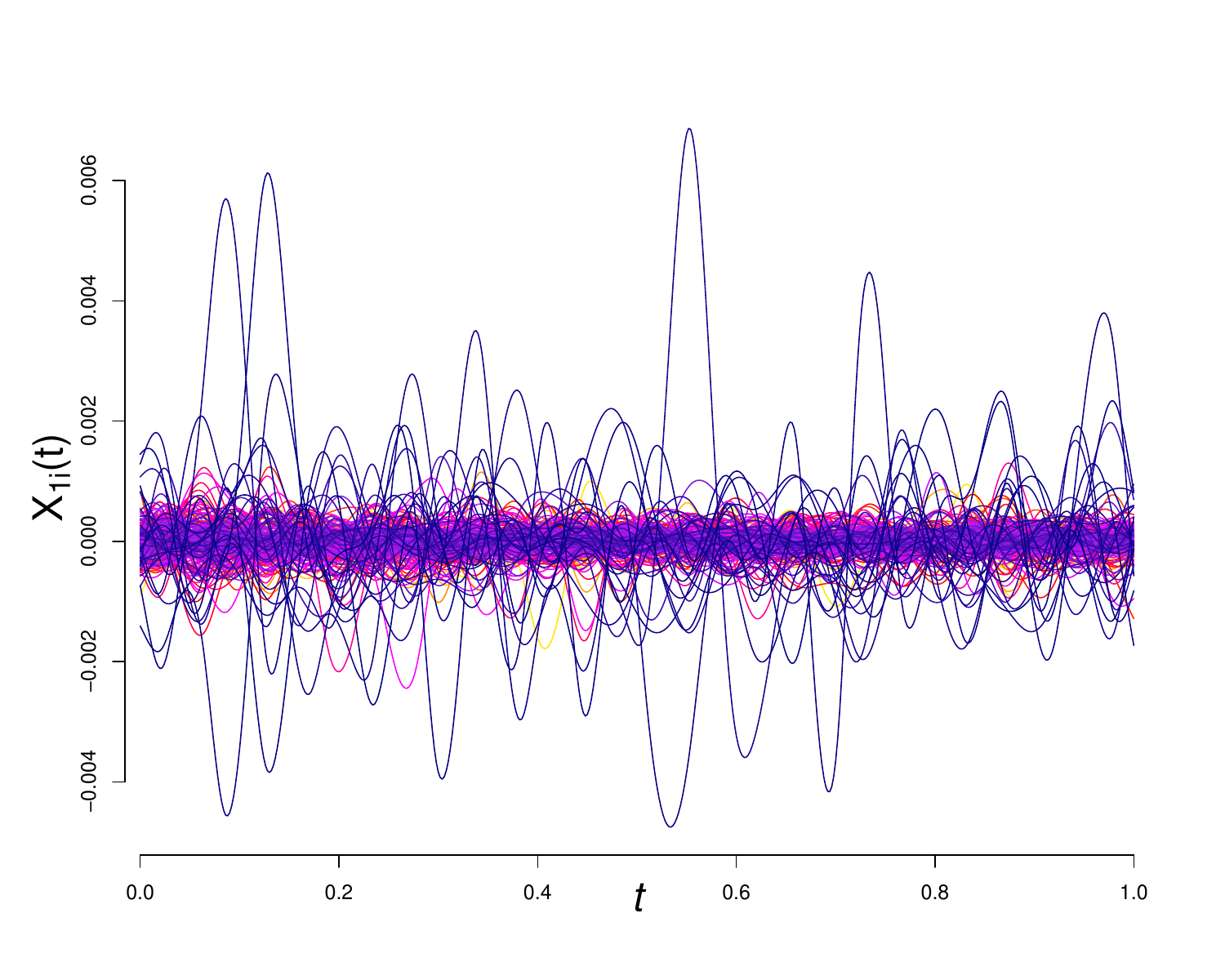}
    \caption*{(a)}
\end{minipage}
\begin{minipage}[c]{.5\textwidth}
    \centering
    \includegraphics[width=\textwidth]{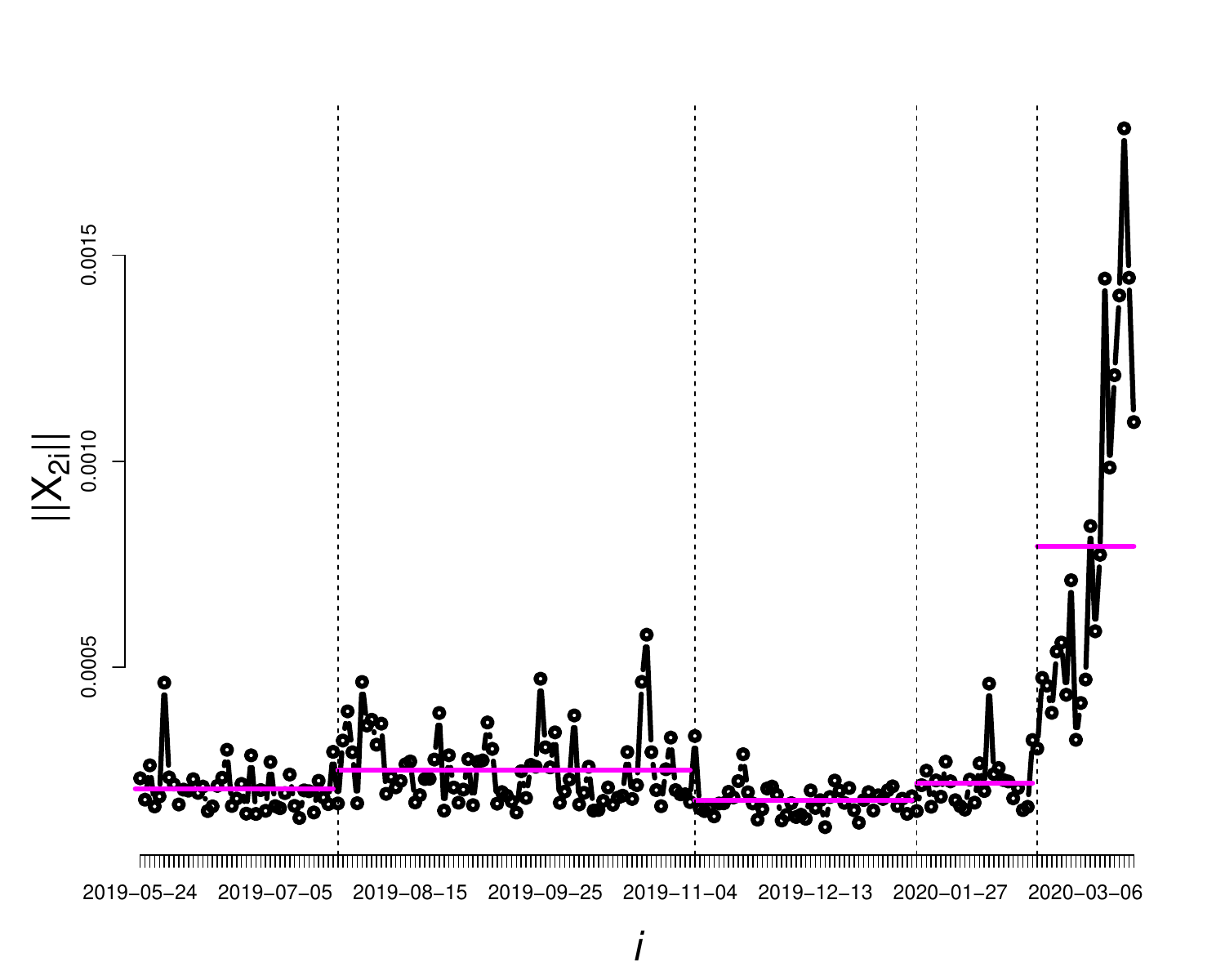}
    \caption*{(b)}
\end{minipage}
    \caption{(a) Twitter differenced log returns and (b) norms of the Twitter differenced log returns over time, with the FKWC detected changepoints and changepoint interval means overlaid.  }
    \label{fig::twtr_returns}
\end{figure}
In this section we present an application of the multiple changepoint FKWC procedure to intraday differenced log returns of \texttt{twtr} stock. 
We analyse 207 daily asset price curves of \texttt{twtr} starting on June 24th 2019 and ending March 20th 2020. 
The price was measured in one minute intervals over the course of the trading day, resulting in a total of 390 minutes per day. 
In order to account for edge effects from smoothing the curves, we trimmed 10\% of the minutes from the beginning of the day and 5\% of the minutes from the end of the day.  
This resulted in 332 minutes of stock prices. 
The differenced log returns are defined as
$$X_{ji}(t)=\ln(Y_{ji\floor{331t}+1})-\ln(Y_{ji\floor{331t}}),$$
for $t\in (0,1]$ and where $Y_{jik}$ is the $j^{th}$ asset price on the $i^{th}$ day at minute $k$. 
The data was fit to a b-spline basis, using 50 basis functions, see \texttt{smooth.basis} in the \texttt{fda} \texttt{R} package. 
These data are shown in Figure \ref{fig::twtr_returns}. 
The assumption of zero mean appears to be satisfied here. 
Notice the outliers, which indicates that this data may require robust inference. 
Obviously, these data are not independent, however, we feel that this will not overtly affect our procedure. 
As long as the intervals between changepoints are big enough, we expect the depth values after a change to also change, even if there is say, $m$-dependence in the data. 
We would expect that $m$-dependence could blur the change for a short period of time and cause the changepoint estimate to be biased.

We ran the PELT algorithm with $\kappa_n=3.74+0.3\sqrt{n}=8.06$, as per Section \ref{sec::sim}. 
The FKWC method using the $\MFHD$ depth identified a changepoint on Jan 15 `20 which the $\RP$ depth did not. 
If we include this changepoint, the algorithm identified four changepoints, as given in Table \ref{table::cps_f}. 

Figure \ref{fig::twtr_returns} displays the norms of the curves over time, with the estimated changepoints added as vertical lines and the means of the norms in each interval overlaid. 
We can see clear changes in the mean of the norms during these periods. 
We may also notice that our procedure was unaffected by the outlier at the beginning of the series and the one just before the last estimated changepoint. Table \ref{table::cps_f} gives the magnitude and sign of the changes as well. 
We can see that the largest changepoint is the last one; clearly attributed to the instability caused by the COVID-19 pandemic. 
It is interesting to see whether or not the other changepoints occurred due to market wide behaviour, or events specific to social media or even just Twitter itself. 
For example, running the same algorithm on \texttt{snap} stock over the same period of time reproduces the changepoints on Nov 07 `19 and Feb 21 `20 but not the other two changepoints. 
One possibility for the estimated changepoint on July 2019 could be the Twitter earnings report released just prior, e.g., \citep{laurenfeiner_2019}. 

Aside from determining possible causes for changepoints, from a modelling perspective, one may wish to avoid using a functional GARCH model. 
This could be due to the fact that in order to fit a functional GARCH model at the present time, one must choose to fit the functional data to a relatively small number of basis functions in order to keep to the number of parameters in the GARCH model small. 
If no clear basis exists, and the principle component analysis does not work well due to outliers, one may wish for an alternate approach. 
Instead one can remove the heteroskedasticity in the data by re-normalizing the curves in each interval, and then proceed with alternative time series modelling from there. 
Of course, this would not estimate future changepoints; one could model the changepoint process and the return curves separately. 
\begin{table}[t]
\centering
\begin{tabular}{cc}
\toprule
Interval &  Centered Rank Mean \\ \midrule
    Jun 24 `19 - Jul 24 `19 & 19.73  \\
    Jul 24 `19- Nov 07 `19 & -15.64 \\
    Nov 11 `19- Jan 15 `20 & 46.04  \\
    Jan 15 `20- Feb 21 `20 & 1.08\\
    Feb 21 `20 - Mar 20 `20 & -90.80
\end{tabular}
\caption{Changepoints and centered $\MFHD$ rank means. Notice the largest change occurs at the last changepoint.}
\label{table::cps_f}
\end{table}

\subsection{Resting state f-MRI pre-processing}
Functional magnetic resonance imaging, or f-MRI, is a type of imaging for brain activity. 
f-MRI uses magnetic fields to determine oxygen levels of blood in the brain in order to produce 3-dimensional images of the brain. 
Many of these images are taken over a period of time, which results in a time series of 3-dimensional images. 
Note that each MRI in a given subject's f-MRI can be viewed as a function on $[0,1]^3$. 
Resting state f-MRI is a type of f-MRI data where no intervention is applied to the subject during the scanning process. 
f-MRI scans go through extensive pre-processing before being analysed. 

One assumption commonly made is that, after several pre-processing steps, each subject's resulting functional time series is stationary. 
It is therefore important to check the scans at an individual level in order to ensure that each time series is stationary. 
For subjects whose time series is not stationary, we must make the necessary corrections or exclusions from the ensuing data analysis. 
The covariance kernel of an f-MRI time series is a 6-dimensional function. 
Existing methods make a separability assumption on the covariance kernel  \citep{Stoehr2019}, which we do not make here. 
Additionally, \cite{Stoehr2019} mentions the need for a robust method of detecting non-stationarities in f-MRI data, which leads us to apply the FKWC procedure to this data. 
We analyse several scans from the Beijing dataset, which were retrieved from \url{www.nitrc.org}. 
These scans were also analysed by \cite{Stoehr2019}. 
Following instruction provided by \url{https://johnmuschelli.com/}, we performed the following pre-processing steps to the data. 
We trimmed the first 10 seconds from the beginning of the scan, in order to have a stable signal. 
We then performed rigid motion correction using \texttt{antsMotionCalculation} function in the \texttt{ANTsR R} package. 
A 0.1 Hz high-pass Butterworth filter of order 2 was applied voxel-wise to remove drift and trend from the data. 
We then removed 15 observations from either end of the time series in order to remove the edge effects of the filter. 
The gradient of each scan was then estimated using the \texttt{numDeriv} package, which resulted in four time series of functional data, where each function is a three-dimensional image.

We then computed the $\RP$ sample depth values as follows. 
First we projected each of the four time series' onto 50 unit functions. 
Then, for each of the 50 projected time series, we computed the halfspace depth values of each four-dimensional observation. 
We then averaged these depths over the 50 unit vectors. 
We use the half-space depth since it is faster to compute than the simplicial depth for four-dimensional data. 
We then applied the FKWC hypothesis tests and the FKWC multiple changepoint algorithm to the resulting depth values. 
In addition, we restricted estimated changepoints to be at least 10 observations away from either boundary. 
Code to run the FKWC procedure on three-dimensional functional data can be retrieved from Github \citep{code}. 

\begin{table}[ht]
\centering
\begin{tabular}{r|rr|rrr}
 \multicolumn{1}{c}{ }  &\multicolumn{2}{c}{AMOC} &  \multicolumn{3}{c}{Epidemic}\\
  \hline
 Subject & Estimate & p-value & Estimate 1 & Estimate 2 & p-value \\ 
  \hline
sub08455 & 116.00 & 0.36 & 34.00 & 54.00 & 0.95 \\ 
  sub08992 & 35.00 & 0.00 & 36.00 & 175.00 & 0.00 \\ 
  sub08816 & 39.00 & 0.20 & 43.00 & 94.00 & 1.00 \\ 
  sub34943 & 159.00 & 0.10 & 21.00 & 173.00 & 0.01 \\ 
  sub12220 & 30.00 & 0.09 & 31.00 & 127.00 & 0.25 \\ 
  sub06880 & 116.00 & 0.00 & 24.00 & 117.00 & 0.00 \\ 
   \hline
\end{tabular}
\caption{Changepoint estimates and p-values resulting from running the FKWC changepoint tests on the different subjects.}
\label{tab::fmri}
\end{table}
\begin{figure}
\begin{minipage}[c]{.32\textwidth}
    \centering
    \includegraphics[width=\textwidth]{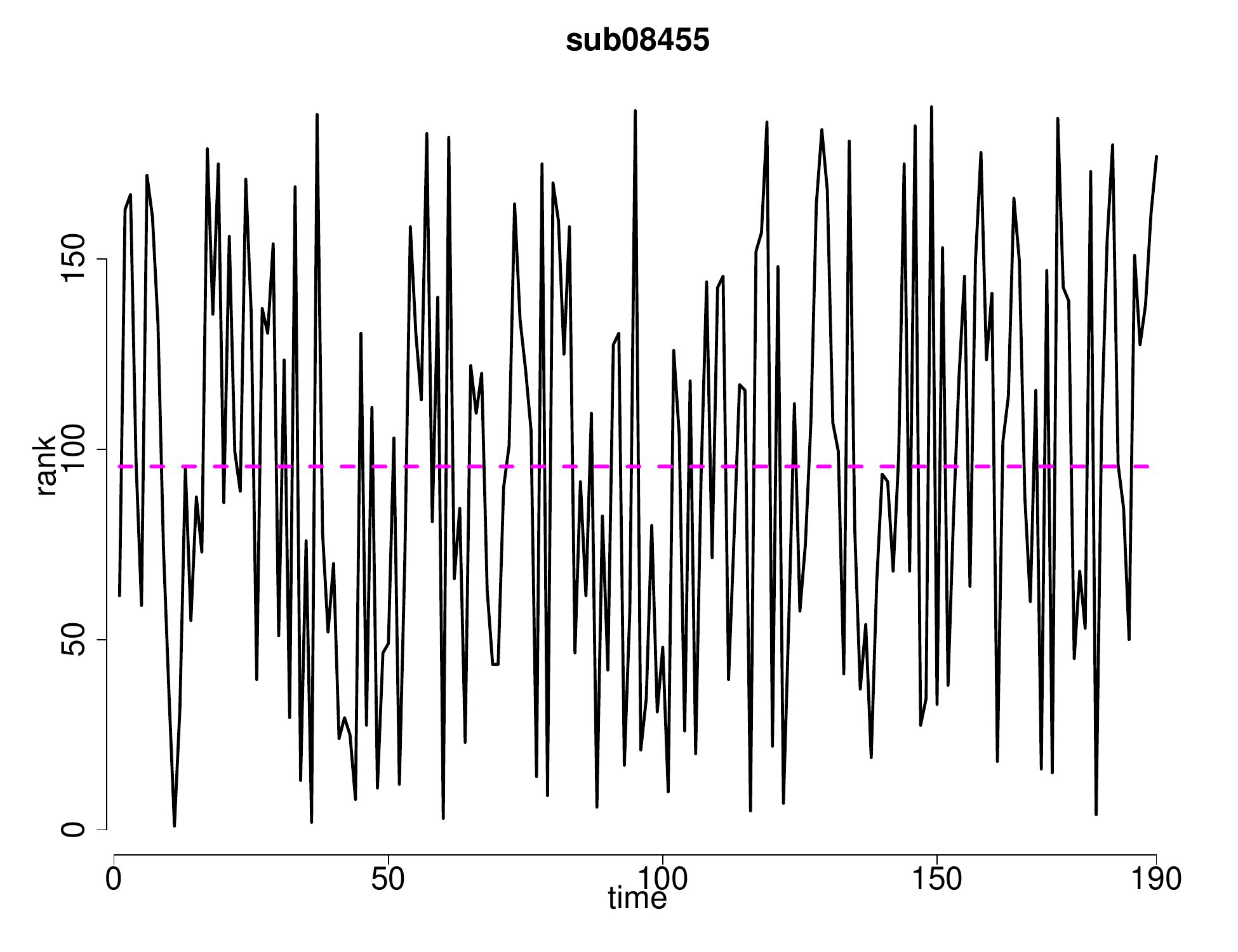}
\end{minipage}
\begin{minipage}[c]{.32\textwidth}
    \centering
    \includegraphics[width=\textwidth]{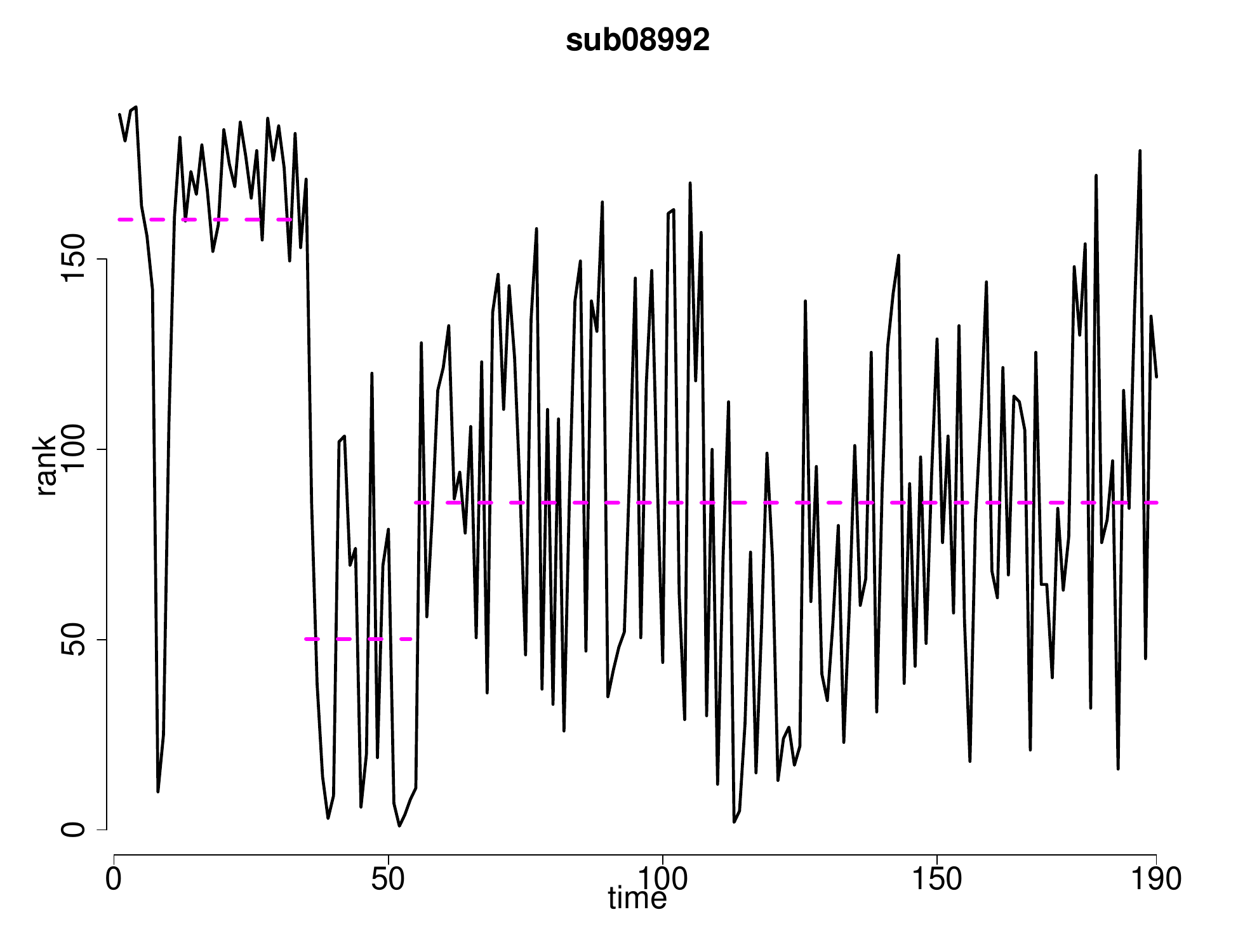}
\end{minipage}
\begin{minipage}[c]{.32\textwidth}
    \centering
    \includegraphics[width=\textwidth]{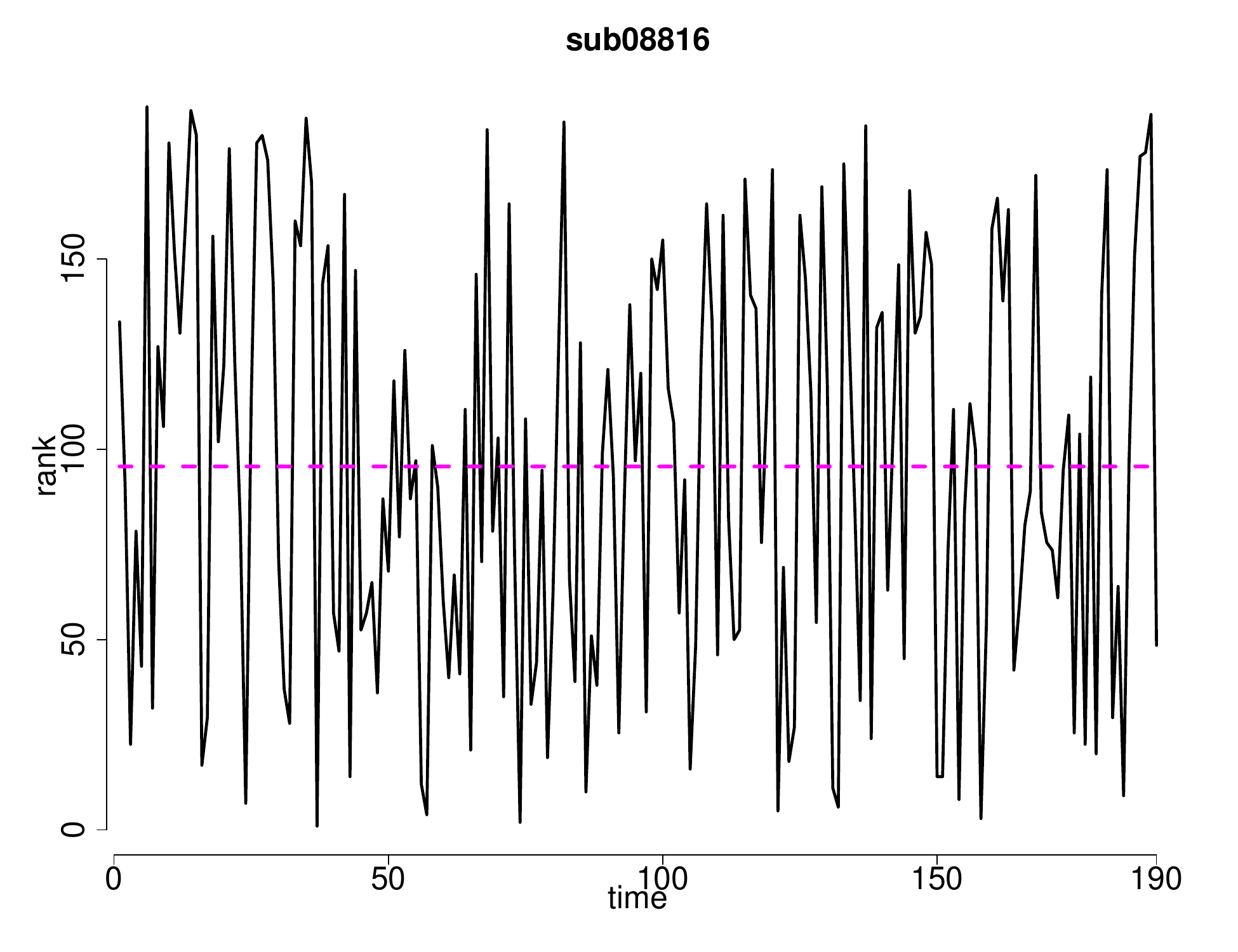}
\end{minipage}\hfill\newline
\begin{minipage}[c]{.32\textwidth}
    \centering
    \includegraphics[width=\textwidth]{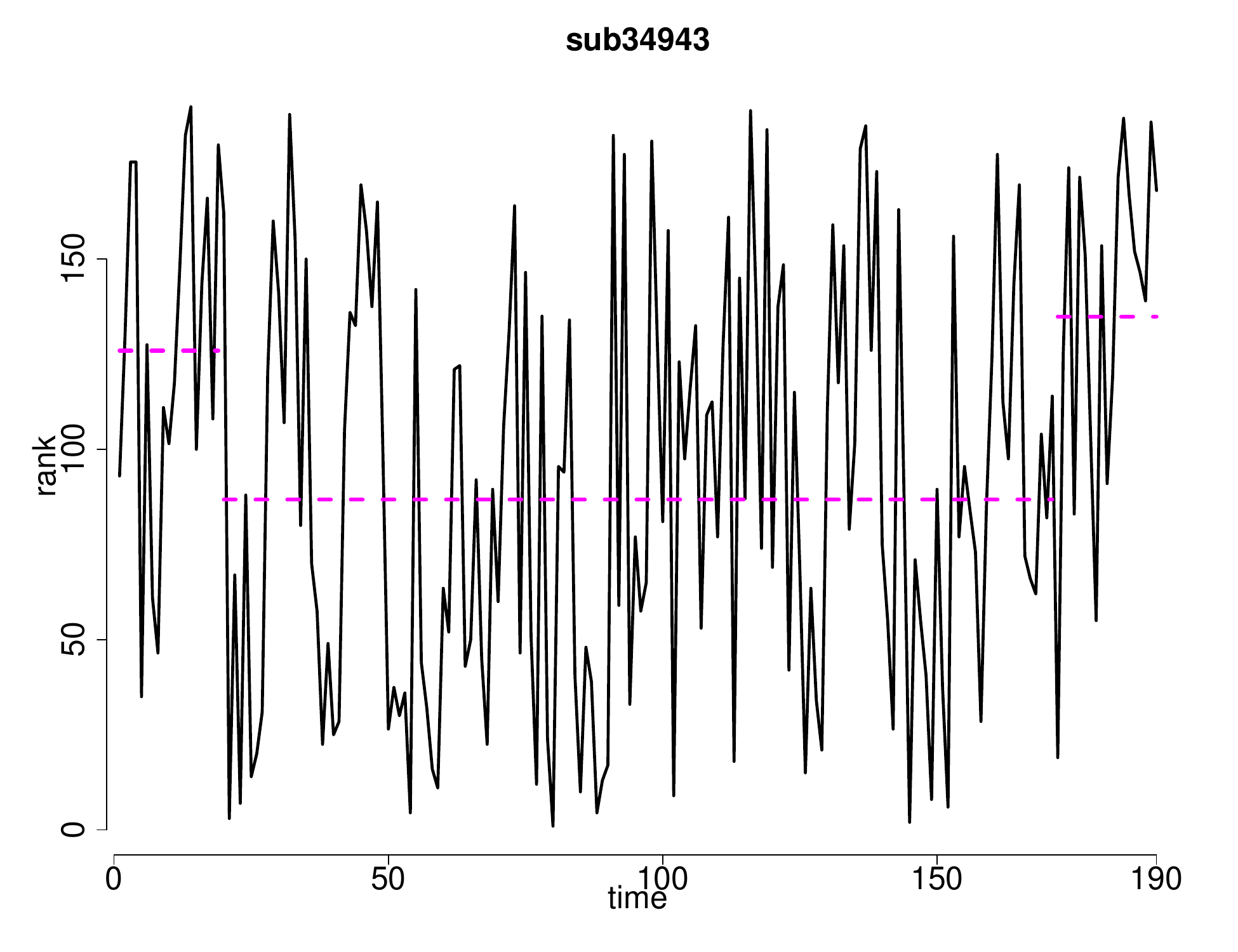}
\end{minipage}
\begin{minipage}[c]{.32\textwidth}
    \centering
    \includegraphics[width=\textwidth]{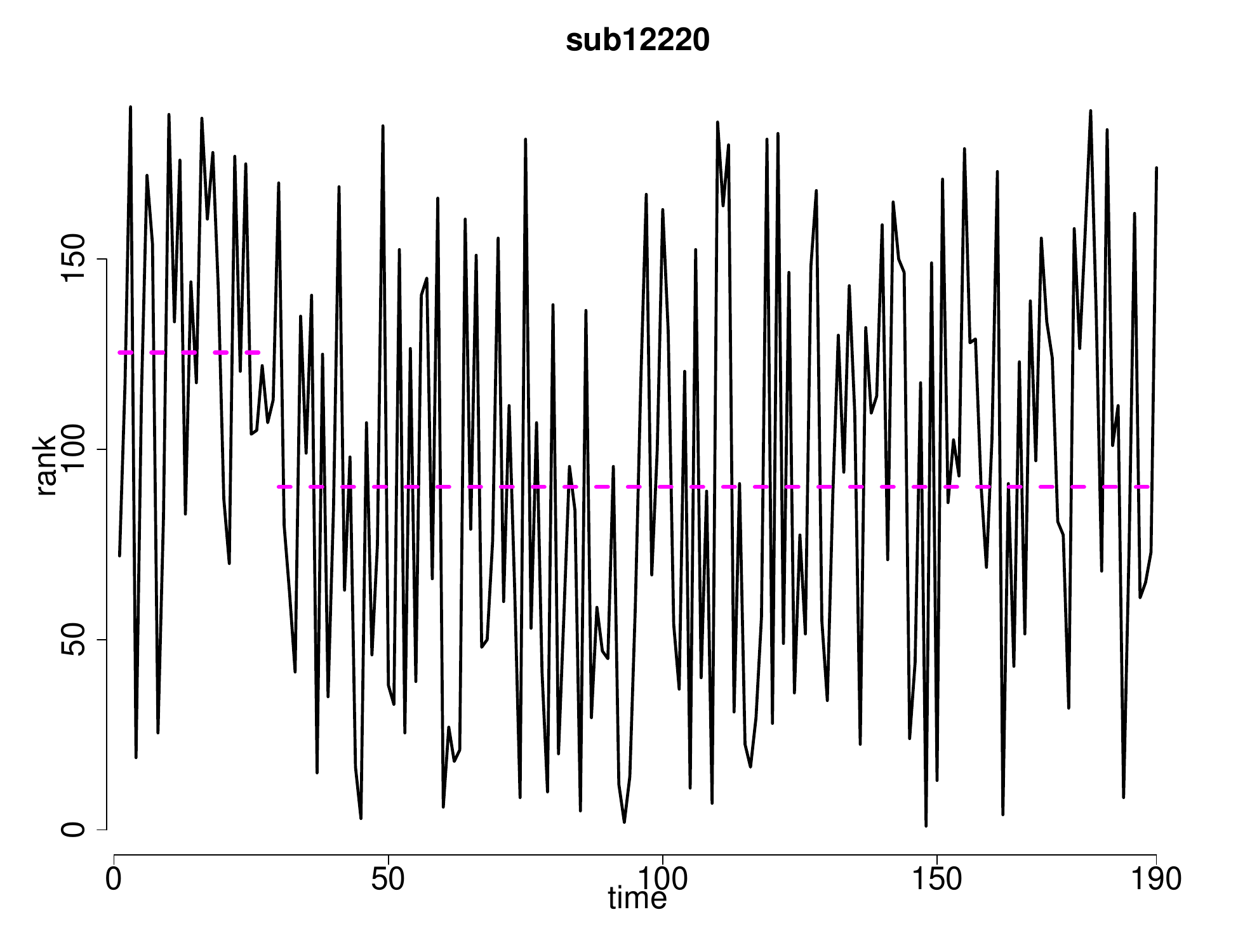}
\end{minipage}
\begin{minipage}[c]{.32\textwidth}
    \centering
    \includegraphics[width=\textwidth]{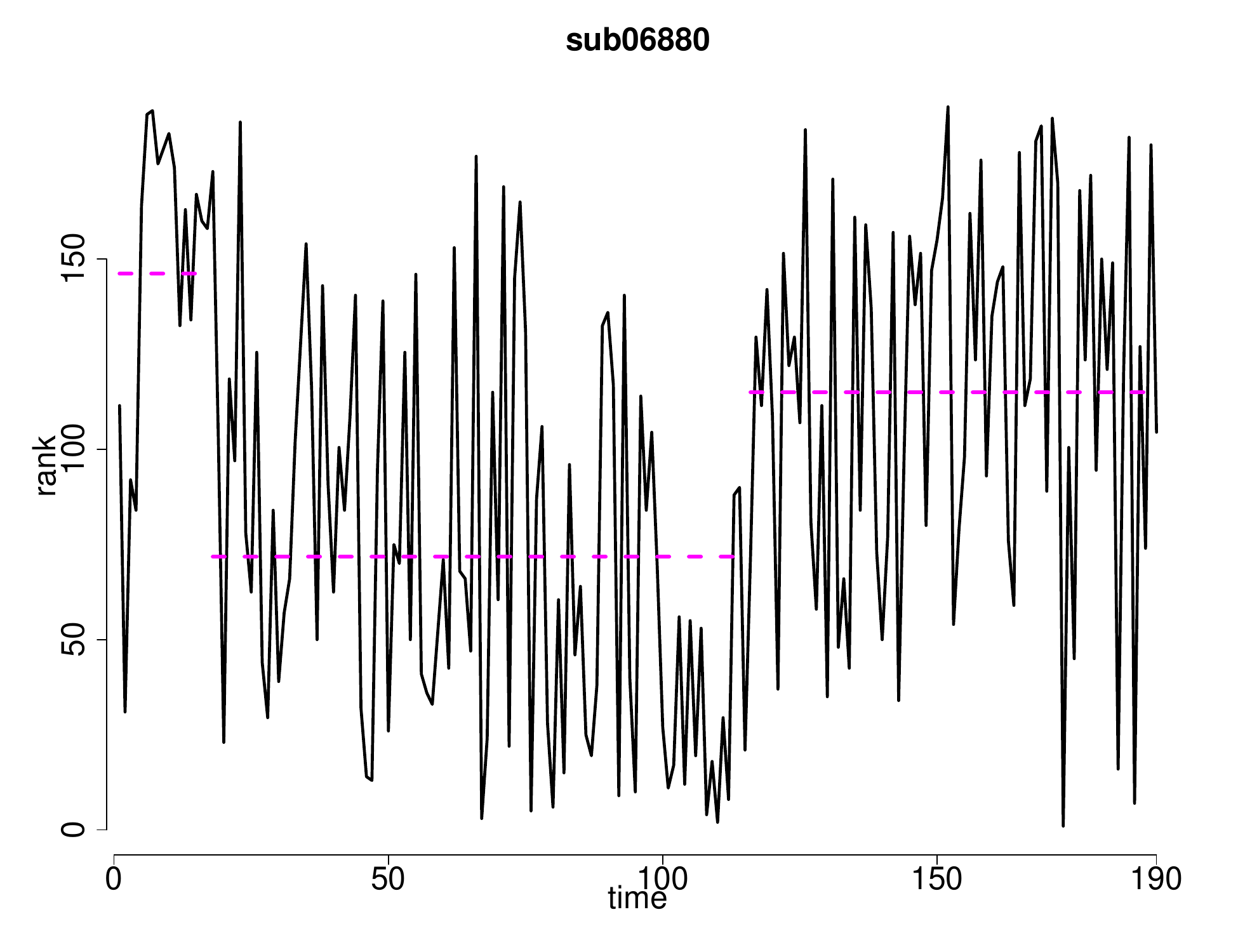}
\end{minipage}
\caption{Ranks of the random projection depth values for several f-MRI scans with detected changepoint means overlaid.}
\label{fig::fmri}
\end{figure}
Figure \ref{fig::fmri} contains the ranks of the random projection depth values for several f-MRI scans, with the resulting changepoint intervals identified by the FKWC multiple changepoint algorithm overlaid. 
Table \ref{tab::fmri} contains the p-values and changepoint estimates resulting from running the hypothesis testing procedures. 
We see that changes are detected in four of the six subjects analysed, two of which appear to be an epidemic type change; the ranks return to their previous means after the second interval. 
This is consistent with the idea that the epidemic model is more suitable for some resting state f-MRI scans \citep{Stoehr2019}. 
For subject \texttt{sub08455}, we do not detect any changes in the sequence, even though this subject's f-MRI has an outlier early in the sequence. 
This outlier can create a false positive for the AMOC alternative, as discussed by \cite{Stoehr2019}. 
Notice that the p-values are not small for this subject when running our test. 
The rank sequence for subject \texttt{sub08992} was estimated to have two changepoints, though the distribution of ranks in the first and third intervals are clearly different. 
This is why the estimates from the epidemic model and the multiple changepoint procedures differ. 
Though, the null hypothesis is rejected by both the AMOC and epidemic model tests, even if we were to use any p-value correction procedure. 
In addition, the FKWC procedure ignores the outlier at the beginning of the sequence for subject \texttt{sub08992}, which showcases the robustness of the FKWC procedure. 
The FKWC procedure did not detect any changepoints for subject \texttt{sub08816}, whereas the methods of \citet{Stoehr2019} detected an epidemic period. 
This could be due to differences in pre-processing, trimming, or the nature of the different methods' assumptions. 
For subject \texttt{sub34943} we see that an epidemic change is detected by the multiple changepoint procedure, and the p-value from the hypothesis is borderline significant with p-value corrections (0.01). 
In the case of the AMOC test, the null hypothesis is not rejected. 
We note that no change was detected by the functional procedure in \citet{Stoehr2019}, but a change was detected in the multivariate procedure. 
For subject \texttt{sub12220}, one change is detected, by the multiple changepoint procedure, though the hypothesis testing yields non-significant results. 
We remark that the test of \cite{Stoehr2019} detects a change. 
The location of the change detected by \cite{Stoehr2019} occurs early in the sequence, and, as a result of the trimming we applied to the sequence, occurs very early in our time series. 
This makes it difficult to detect by our procedure. 
For subject \texttt{sub06880} we see that all three components of the FKWC procedure agree that there are changepoints. \citet{Stoehr2019} also detected changepoints in this subject's sequence of observations. 
\section*{Funding acknowledgement}
The authors acknowledge the support of the Natural Sciences and Engineering Research Council of Canada (NSERC). Cette recherche a \'et\'e financ\'ee par le Conseil de recherches en sciences naturelles et en g\'enie du Canada (CRSNG),  [DGECR-2023-00311].
\bibliography{FCP_COV_PAPERS}
\bibliographystyle{apalike}
\hfill\newpage
\appendix
\section{Comments on multivariate functional depth functions}\label{app::depth_al}
This section contains additional details on multivariate functional depth functions.
\subsection{Explicit definitions of multivariate functional depth functions}\label{app::depth}
We consider the random projection depth and the multivariate halfspace depth. 
We first consider the random projection depth. 
Here, define $\nu_u$ to be the law of $(\ip{X_1,u},\ldots, \ip{X_p,u})$ if $X\sim \nu$ and let $S=\{u\in\ltwo\colon\ \norm{u}=1\}$, where $\norm{\cdot}$ is the $\ltwo$-norm with respect to the Lebesgue measure on $[0,1]^d$.  
As stated in Section \ref{sec::depth}, the random projection depth takes $\Omega=S$, $\Dd$ any multivariate depth function, $Q\in \cM_1(S)$ and $g(x,u)=(\ip{x_1,u},\ldots,\ip{x_p,u})$. 
In simulation, we take $\Dd$ to be the simplicial depth \citep{Liu1988}. 
The Simplicial depth is $(p+1,\cF)$-regular with $\VC(\cF)=p^2\log(p)$ \citep{ramsay2022concentration}. 
Let $\Delta(x_1, \ldots,x_{d+1})$ be the $d$-dimensional simplex with vertices $x_1, \ldots,x_{d+1}$. 
Suppose that $Y_1, \ldots,Y_{d+1}$ are i.i.d.\ from some $\mu~\in~\cM_1(\rdd)$. The simplicial depth of a point $x\in \rdd$ with respect to $\mu$ is
\begin{equation*}
    \SD (x,\mu)= \Pr\left(x\in \Delta(Y_1, \ldots,Y_{d+1})\right).
\end{equation*}
The random projection depth \citep{Cuevas2007} is defined as follows: 
\begin{definition}\label{def::rpd}
Let $u_1,\ldots,u_M\sim \mu\in\cM_1(S)$. 
The random projection depth, with base probability measure $\mu$, of a $p$-dimensional vector of functions $x\in(\ltwo)^p$ with respect to $\nu\in\cM_1((\ltwo)^p)$ is
$$\RP(x,M,\nu,\mu)=\frac{1}{M}\sum_{m=1}^M\SD(\ip{x,u_m},\nu_{u_m}).$$
\end{definition}
The random projection depth of a point $x$ is the average projected, multivariate depth of $x$ with respect to some base probability measure $\mu$. 
Here, we take $\mu$ to be such that $u_1,\ldots,u_M$ are zero mean Gaussian processes with exponential variogram $\gamma(s,t)=\exp(-5|s-t|)$, standardized such that they have unit norm. 

Define the multivariate halfspace depth of a point $x\in\re^p$ with respect to $\mu\in\cM_1(\rdd)$ as 
$$\HD(x,\mu)=\inf_{u\in\re^p,\ \norm{u}=1}\mu(X^\top u\leq x^\top u ).$$
Let $\nu_t$ be the law of $X(t)$ if $X\sim \nu\in \cM_1((\ltwo)^p)$.
The multivariate functional halfspace depth is defined as follows. 
\begin{definition}\label{def::mfhd}
The multivariate functional halfspace depth of a $p$-dimensional vector of continuous functions $x\in(\ltwo)^p$ with respect to $\nu\in\cM_1((\ltwo)^p)$ is
\begin{equation*}
\MFHD(x ,\nu)=\int_{[0,1]^d} \HD(x(t),\nu_t)dt.
\end{equation*}
\end{definition}
The multivariate functional halfspace depth is the average over $[0,1]^d$ of the multivariate depth of $x(t)$ with respect to $\nu_t$. 
Note that the multivariate functional halfspace depth was introduced in a more general form \citep{slaets11, Hubert2012, Claeskens2014}. 
Furthermore, when $p=1$, this depth function is also equivalent to the depth of \citet{Fraiman2001}. 
As in the case of the random projection depth, one can replace the halfspace depth in Definition \ref{def::mfhd} with another multivariate depth function, such as simplicial depth \citep{Fraiman2001}. 

\subsection{Connections between depth and covariance}\label{app::depth_cov}
We provide a brief summary of the connection between the distribution of $\ID(X_{k_i+1},\nu)-\ID(X_{k_i},\nu)$ and the covariance operators $\kerrO_i$, $\kerrO_{i+1}$. 
To this end, we must first define the \emph{symmetry weight function}. 
Consider some univariate probability measure $P\in\cM(\re)$ which possesses the property $\int_{\re} Xd\mu=0$. 
Suppose that $P$ also possesses a twice differentiable cumulative distribution function $F_P$, by which we denote the first and second derivative of $F_P$ by $f_P$ and $f_P'$, respectively.
The symmetry weight function of $P$ is 
$$\SW(P)= \left(\frac{1}{2}-F_P(0)\right)f_P'(0)+f_P(0)^2.$$ 
The first term is a signed measure of skew, whereas the second term is large when there is high density at the mean. 
For the $\RP$ depth function, as defined in Appendix \ref{app::depth}, the connection can be summarized as follows: 
Suppose that $\nu$ is such that
\begin{equation}\label{eqn::mm}
    \E{}{\ID(X_{k_1+1},\nu)-\ID(X_{k_1},\nu)}\neq 0 \implies \med(\ID(X_{k_1+1},\nu)-\ID(X_{k_1},\nu))\neq 0.
\end{equation}
Then if $\ID=\RP$, $\SW(\nu_u)$ is well-defined for all $u\in S$ and $\E{\nu}{\norm{X}^3}<\infty,$ a sufficient condition for the jump sizes to be non-zero, i.e., $\Prr{\ID(X_{k_1},\nu)\leq \ID(X_{k_1+1},\nu)}\neq 1/2$, is
\begin{equation*}   
    \int_{S}\SW(\nu_u) \ip{\kerrO_2 u,u }dQ+\mathcal{R}_1\neq \int_S\SW(\nu_u)\ip{\kerrO_{1} u,u }dQ+\mathcal{R}_2.
\end{equation*}
Here $\mathcal{R}_1,\mathcal{R}_2$ are remainder terms which are expected to be small see \citep{Ramsay2023b}. 
Then, if $\ID=\MFHD$ and \eqref{eqn::mm} holds, a sufficient condition for $\Prr{\ID(X_{k_1},\nu)\leq \ID(X_{k_1+1},\nu)}\neq 1/2$ to hold is
\begin{equation*}
\int_{[0,1]^d} \Ee{}{|1/2-F_{\nu_t}(X_{k_1}(t))|}dt\neq \int_{[0,1]^d} \Ee{}{|1/2-F_{\nu_t}(X_{k_1+1}(t))|}dt.
\end{equation*}
Now, assuming $F_{\nu_t}$ is twice differentiable and the following notion of symmetry $X\eqd -X$ for $X\sim\nu$. 
Taking a first order Taylor expansion of $F_{\nu_t}$ yields
\begin{equation}\label{eqn::changetype_4}
\int_{[0,1]^d} \Ee{}{|f_{\mu_t}(0)X_{k_1}(t)+\mathcal{R}_3(t)|}dt\neq \int_{[0,1]^d} \Ee{}{|f_{\mu_t}(0)X_{k_1+1}(t)+\mathcal{R}_4(t)|}dt,
\end{equation}
where $\mathcal{R}_3,\mathcal{R}_4$ are the remainder terms resulting from the second order Taylor expansion of $F_{\nu_t}$. 
Notice that the quantity \eqref{eqn::changetype_4} is roughly a functional data analogue of the median absolute deviation. 

\section{Proofs}\label{sec::Proofs}
We now prove Theorem~\ref{thm::main-result} with a series of lemmas. 
We first prove a bound on the expectation of $\sup_{x\in(\ltwo)^p}|\ID(x,\hat\nu)-\ID(x,\nu)|$. 
Let $\nu=\sum_{i=1}^{\ell+1}(k_i-k_{i-1})\nu^i/n$. 
\begin{lemma}\label{lem::exp_bound}
If Conditions \ref{cond::contin}--\ref{cond::depth} hold, then, for all $d,p,n\geq 1$ we have that
$$\Eeee\sup_{x\in(\ltwo)^p}|\ID(x,\hat\nu)-\ID(x,\nu)|\lesssim K \ell\sqrt{\VC(\sF)/n}.$$
\end{lemma} 
\begin{proof}
Take $U\sim Q$ such that $U$ is independent of $\{X_i\}_{i=1}^n$. Let $\Eeee_{\mathbf{X}}$ denote the expectation with respect to $\{X_i\}_{i=1}^n$ and let $\Eeee_{U}$ denote the expectation with respect to $U$. 
Then, we have that by Jensen's inequality
\begin{align*}
    \Eeee\sup_{x\in\ltwo[0,1]}|\ID(x,\hat\nu)-\ID(x,\nu)|&=\Eeee\sup_{x\in\ltwo[0,1]}\left|\int_{\Omega} \Dd(g(x,u),\hat\nu_u) dQ(u)-\int_{\Omega} \Dd(g(x,u),\nu_u) dQ(u)\right|\nonumber\\
&\leq \Eeee\int_{\Omega} \sup_{y\in\re^{p}}|\Dd(y,\hat\nu_u)- \Dd(y,\nu_u)|dQ(u) 
\end{align*}
Now, using Condition \ref{cond::depth} and the Fubini–Tonelli theorem:
\begin{equation}\label{eqn::upper_bound_diff}
    \Eeee\int_{\Omega} \sup_{y\in\re^{p}}|\Dd(y,\hat\nu_u)- \Dd(y,\nu_u)|dQ(u)\leq \Eeee_{\mathbf{X}}\Eeee_{U} Kd_{\cF}(\hat\nu_U,\nu_U)=K\cdot\Eeee_{U} \Eeee_{\mathbf{X}}( d_{\cF}(\hat\nu_U,\nu_U)).
\end{equation}
Lastly, applying equation (85) of \citep{sen2022gentle} and the triangle inequality, 
\begin{align*}
K\cdot\Eeee_{U} \Eeee_{\mathbf{X}}( d_{\cF}(\hat\nu_U,\nu_U))&\lesssim  \ell K\cdot\Eeee_{U}  \sqrt{\VC(\sF)/n}
= \ell K\sqrt{\VC(\sF)/n}. \qedhere
\end{align*}
\end{proof}
We now prove a concentration inequality for the integrated depths. 
\begin{lemma}\label{lem::cb}
If Conditions \ref{cond::contin}--\ref{cond::depth} then, there exists universal constant $c,c'>0$ such that, for all $d,p,n\geq 1$, it holds that
\begin{equation*}
    \Prr{\sup_{x\in(\ltwo)^p}|\ID(x,\hat\nu)-\ID(x,\nu)|\geq t+cK\ell\sqrt{\VC(\sF)/n}}\leq 2\exp\left(c'\frac{-nt^2}{K\ell\sqrt{\VC(\cF)/n}+1}\right).
\end{equation*}
\end{lemma}
\begin{proof}
Directly from \eqref{eqn::upper_bound_diff}, we have that 
\begin{align*}
\sup_{x\in(\ltwo)^p}|\ID(x,\hat\nu)-\ID(x,\nu)|&\leq K\cdot\Eeee_U( d_{\cF}(\hat\nu_U,\nu_U))=K\cdot\Eeee_U\sup_{f\in\cG}\left|\frac{1}{n}\sumn f(X_i,U)\right|,
\end{align*}
where 
$$\cG=\left\{h\colon h(x,u)=f(g(x,u))-\int f(g(x,u))d\nu_u(x),\ f\in\cF\right\}.$$
Next, for all $s>0$, we have that
\begin{align*}
    \Prr{\Eeee_U\sup_{f\in\cG}\left|\frac{1}{n}\sumn f(X_i,U)\right|\geq s}&\leq \Prr{\Eeee_U\sup_{f\in\cG}\frac{1}{n}\sumn f(X_i,U)\geq s}+\Prr{\Eeee_U\sup_{f\in\cG}\frac{1}{n}\sumn -f(X_i,U)\geq s}\\
    &\leq \Prr{\sup_{f\in\cG}\frac{1}{n}\sumn \Eeee_Uf(X_i,U)\geq s}+\Prr{\sup_{f\in\cG}\frac{1}{n}\sumn -\Eeee_Uf(X_i,U)\geq s}\\
    &\leq 2 \Prr{\sup_{f\in\cG'}\frac{1}{n}\sumn f(X_i)\geq s},
\end{align*}
where $$\cG'=\left\{\pm h\colon\ \ h(x,u)=\int_\Omega \left[f(g(x,u))-\int_{\re^p} f(g(x,u))d\nu_u\right]dQ,\ f\in\cF\right\}.$$
Now, observe the following three facts:
\begin{enumerate}
    \item For any $f\in\cG'$ it holds that $\int_{(\ltwo)^p} f(x)d\nu=0$.
    \item A direct application of Lemma \ref{lem::exp_bound} gives that $\Eeee\sup_{f\in\cG'}\frac{1}{n}\sumn f(X_i)\lesssim K\ell\sqrt{\VC(\cF)/n}$. 
    \item By assumption, $\sup_{f\in\cG'}\norm{f}_\infty\leq 1$. 
\end{enumerate}
Letting $Z=\sup_{f\in\cG'}\frac{1}{n}\sumn f(X_i)$, the above three facts allow us to apply Talagrand's concentration inequality for empirical processes \citep{talagrand1996new, bousquet2003concentration}. 
This application yields that there exists a universal constant $c>0$ such that
\begin{align*}
   \Prr{Z\geq t+\Eeee Z}&\leq\Prr{Z\geq t+cK\ell\sqrt{\VC(\cF)/n}}\leq \exp\left(\frac{-nt^2}{4cK\ell\sqrt{\VC(\cF)/n}+2+2t/3}\right).
\end{align*}
Next, using the fact that $t\in [0,1]$, we have that there exists a universal constant $c'>0$ such that 
\begin{align*}
   \Prr{Z\geq t+cK\ell\sqrt{\VC(\cF)/n}}\leq \exp\left(\frac{-nt^2}{4cK\ell\sqrt{\VC(\cF)/n}+5}\right)\leq \exp\left(c'\frac{-nt^2}{K\ell\sqrt{\VC(\cF)/n}+1}\right)
\end{align*}
This completes the proof. 
\end{proof}
Next, we prove that the rank means based on the sample depth function concentrate around those based on the population depth function. 
For $1\leq s<e\leq n$, let $n_{s,e}=e-s+1$, $$\overline{\widehat{R}}_{s,e}=\frac{1}{e-s+1}\sum_{i=s}^e \widehat{R}_i\qquad\text{and}\qquad \overline{R}_{s,e}=\frac{1}{e-s+1}\sum_{i=s}^e R_i.$$
In addition, let $\sigma_n^2=n(n+1)/12$.  
Lastly, let $$A_1(c)=\left\{ \max_{1\leq s<e\leq n} |\frac{\overline{\widehat{R}}_{s,e}-\overline{R}_{s,e}}{\sigma_n}|\geq \sqrt{\frac{c\lambda_{n,\cF,K}}{n_{s,e}}}\right\}\qquad\text{and}\qquad A_2(c)=\left\{ \max_{1\leq s<e\leq n} |\frac{\overline{R}_{s,e}-\Eeee\overline{R}_{s,e}}{\sigma_n}|\geq \sqrt{\frac{c\lambda_{n,\cF,K}}{n_{s,e}}}\right\}.$$
\begin{lemma}\label{lem::rank-con}
If Conditions \ref{cond::contin}--\ref{cond::depth} hold then there exists a universal constant $C>0$ such that
$\Prr{A_1(C)\cup A_2(C)}\lesssim 1/n$.
\end{lemma}
\begin{proof}
It suffices to show that $\Prr{A_1(C)}\lesssim 1/n$ and $\Prr{A_2(C)}\lesssim 1/n$. 
We begin by showing that $\Prr{A_1(C)}\lesssim 1/n$. 
Making use of Condition~\ref{cond::depth}, item 2.\ and equations (A3) and (A4) from \citet{Chenouri2020DD}, we have that 
\begin{align}
    \max_{1\leq s<e\leq n} |\sqrt{n_{s,e}}\frac{\overline{\widehat{R}}_{s,e}-\overline{R}_{s,e}}{\sigma_n}|&\lesssim  \max_{1\leq s<e\leq n}\frac{1}{\sqrt{n_{s,e}}}\sum_{i=s}^{e}  \left|\frac{R_{i}-\widehat{R}_{i}}{\sqrt{\left(n_{s,e}^{2}-1\right) / 12}}\right|\nonumber \\
    \label{eqn::rank-depths-in}
    &\lesssim \max_{1\leq s<e\leq n}K\sqrt{n_{s,e}}\sup_{x\in(\ltwo)^p}\left|\ID(x,\hat\nu_{s,e})-\ID(x,\nu_{s,e})\right|.
\end{align}
Now, using Lemma \ref{lem::cb}, we have that 
$$\Prr{\sup_{x}\left|\ID(x,\hat\nu_{s,e})-\ID(x,\nu_{s,e})\right|\geq t}\leq 2\exp\left(-c'\frac{n_{s,e}\left(t-cK\ell\sqrt{\VC(\cF)/n_{s,e}}\right)^2}{1+K\ell\sqrt{\VC(\cF)/n_{s,e}}} \right),$$
provided $t\gtrsim cK\ell\sqrt{\VC(\cF)/n_{s,e}}$. 
This fact, combined with \eqref{eqn::rank-depths-in} and a union bound, yields that
\begin{align*}
   \Prr{A_1(C)} &\lesssim  n^2\max_{1\leq s<e\leq n}\Prr{\sqrt{n_{s,e}}|\frac{\overline{\widehat{R}}_{s,e}-\overline{R}_{s,e}}{\sigma_n}|\geq \sqrt{C\lambda_{n,\cF,K}}/K}\\
   &\lesssim n^2\exp\left(-c'\frac{\left(C\lambda_{n,\cF,K}-cK\ell\sqrt{\VC(\cF)}\right)}{K^2+K^3\ell\sqrt{\VC(\cF)}} \right).
\end{align*}
Now, taking $C$ large enough yields that 
\begin{align*}
   n^2\exp\left(-c'\frac{\left(C\lambda_{n,\cF,K}-cK\ell\sqrt{\VC(\cF)}\right)}{K^2+K^3\ell\sqrt{\VC(\cF)}} \right)&\lesssim 1/n.
\end{align*}
It remains to show that $\Prr{A_2(C)}\lesssim 1/n$. 
A direct application of Hoeffding's concentration inequality for $U$-statistics \citep{Hoeffding1963} yields that there exists a universal constant $c''>0$ such that
\begin{align*}
   \Prr{A_2(C)} &\lesssim  n^2\max_{1\leq s<e\leq n}\Prr{\sqrt{n_{s,e}}|\frac{\overline{R}_{s,e}-\Eeee\overline{R}_{s,e}}{\sigma_n}|\geq \sqrt{C\lambda_{n,\cF,K}}/K}\lesssim n^2\exp\left(-C\cdot c''\log n\right).
\end{align*}
Again, taking $C$ to be large enough yields that 
$n^2\exp\left(-C\cdot c''\log n\right)\lesssim 1/n$. This completes the proof. 
\end{proof}
\noindent We now prove that the objective function concentrates around its population counterpart. 
Let
$$f(\bfr)=\frac{(r_{j}-r_{j-1})}{\sigma_n^2}\sum_{j=1}^{|\bfr|+1}\left(\Eeee\overline{R}_{j}-\frac{n+1}{2}\right),$$
where the dependence of $\overline{R}_{j}=\sum_{i=r_{j-1}+1}^{r_{j}}R_i/(r_{j}-r_{j-1})$ on $\bfr$ is omitted for brevity. 
Let $\mathbf{K}_n$ denote the set of all ($2^{n-1}$) possible changepoint sets. 
\begin{lemma}\label{lem::KW-Concentrates}
There exists a universal constant $c>0$ such that for all $n,p,d\geq 1$ it holds 
$$\Prr{\bigcup_{\bfr\in\mathbf{K}_n}\left\{|\widehat\cW(\bfr)-f(\bfr)|\lesssim C|\bfr|\lambda_{n,\cF,K}\right\}\big | A_1(C)^c\cap A_2(C)^c}\geq 1-c/n,$$
where $C$ is defined in Lemma \ref{lem::rank-con}.
\end{lemma}
\begin{proof}
In view of the triangle inequality, it suffices to show that 
\begin{equation}\label{eqn::con-goal-1}
    |\widehat\cW(\bfr)-\cW(\bfr)|\lesssim|\bfr|\lambda_{n,\cF,K}
\end{equation}
and 
\begin{equation}\label{eqn::con-goal-2}
   |\cW(\bfr)-f(\bfr)| \lesssim|\bfr|\lambda_{n,\cF,K}.
\end{equation}
We first show \eqref{eqn::con-goal-1}. 
Now, elementary algebra yields that
\begin{align*}
|\widehat\cW(\bfr)-\cW(\bfr)|&\lesssim\left|\sigma^{-2}_{n}\sum_{j=1}^{|\bfr|+1}(r_{j}-r_{j-1})\left(\overline{\widehat{R}}_{j}-\overline{R}_{j}\right)\left(\overline{R}_{j}-\frac{n+1}{2}\right)+\sigma^{-2}_{n}\sum_{j=1}^{|\bfr|+1}(r_{j}-r_{j-1})\left(\overline{\widehat{R}}_{j}-\overline{R}_{j}\right)^2\right|.
\end{align*}
Using this fact yields that, conditional on $A_1(C)^c\cap A_2(C)^c$, we have 
\begin{align*}
|\widehat\cW(\bfr)-\cW(\bfr)|&\lesssim \left|\lambda_{n,\cF,K}^{1/2}\sigma^{-1}_{n}\sum_{j=1}^{|\bfr|+1}\sqrt{(r_{j}-r_{j-1})}\left(\overline{R}_{j}-\frac{n+1}{2}\right)+|\bfr|\lambda_{n,\cF,K}\right|\\
&=\left||\bfr|\lambda_{n,\cF,K}+\lambda_{n,\cF,K}^{1/2}\sigma^{-1}_{n}\sum_{j=1}^{|\bfr|+1}\sqrt{(r_{j}-r_{j-1})}\left[\left(\overline{R}_{j}-\Eeee\overline{R}_{j}\right)+(\frac{n+1}{2}-\Eeee\overline{R}_{j})\right]\right|\\
&\lesssim\left||\bfr|\lambda_{n,\cF,K}+|\bfr|\lambda_{n,\cF,K}+\lambda_{n,\cF,K}^{1/2}\sigma^{-1}_{n}\sum_{j=1}^{|\bfr|+1}\sqrt{(r_{j}-r_{j-1})}\left(\frac{n+1}{2}-
\Eeee\overline{R}_{j}\right)\right|.
\end{align*}
Now, observe that $$\sum_{j=1}^{|\bfr|+1}\sqrt{(r_{j}-r_{j-1})}\left(\frac{n+1}{2}-
\Eeee\overline{R}_{j}\right)=0,$$
which, when combined with the previous inequality, yields \eqref{eqn::con-goal-1}. 
An analogous argument yields \eqref{eqn::con-goal-2}, which completes the proof. 
\end{proof}
We can now prove Theorem~\ref{thm::main-result}. 
\begin{proof}[Proof of Theorem~\ref{thm::main-result}]
We take a contradiction approach, in the spirits of \citet{Wang2019} and \citet{RAMSAY2023}. 
However, our proof relies on Lemmas \ref{lem::cb}--\ref{lem::KW-Concentrates}, as well as some arguments concerning rank statistics, which are novel. 

We first show that the event $\Pr(\hat{\ell}<\ell)\to 0$ as $n\to \infty$.  
To this end, there is at least one changepoint $0<k_{i^*}<n$ such that for any $ j\in [\hat{\ell}]$ it holds that $|k_{i^*}-\hat{k}_j|\geq \Delta_n/2$. 
Now, define $$\bfr_1=\{k_{i^*}-\Delta_n/2,k_{i^*}+\Delta_n/2\}\cup \bfk\backslash k_{i^*}\qquad \text{and}\qquad \bfr_2=\bfr_1\cup \hat\bfk.$$ 
First, observe that $f(\bfr_2)\geq f(\hat\bfk)$, since for any $\bfk_1,\bfk_2\in\mathbf{K}_n$, with $\bfk_1\subset\bfk_2$, we have that $f(\bfk_2)\geq f(\bfk_1)$. 
Now, applying Lemma \ref{lem::KW-Concentrates} in conjunction with the assumed bounds on $\kappa_n$, yields that on $A_1(C)^c\cap A_2(C)^c$, it holds that there exists a universal constant $c>0$ such that
\begin{equation}\label{eqn::step-1-event-1}
      \cW(\bfk)- \cW(\hat\bfk)= f(\bfk)- f(\hat\bfk)-c\ell\kappa_n\geq f(\bfk)- f(\bfr_2)-c\ell\kappa_n.
\end{equation}
The next step is to show that $f(\bfr_2)=f(\bfr_1)$.

At this point, it is helpful to simplify $f(\bfr)$. 
Let $X\sim \nu^i$, $Y\sim \nu^j$ and $p_{ij}=\Prr{\Dd(X,\nu)\geq \Dd(Y,\nu)}$. 
For $i\in[\ell+1]$, $n_i$ be the number of observations coming from each of $\nu^i$. 
With this notation, for any $i\in[\ell+1]$, we have that for $k_{i-1}<R_i\leq k_i$, we have that $ \Eeee{R_i}=\sum_{j=1}^{\ell+1}n_ip_{ij}+1/2$. 
This implies that 
$\Eeee{R_i}-(n+1)/2=\sum_{j=1}^{\ell+1}n_j(p_{ij}-1/2).$, which results in the following statement. 
For any $i\in[\ell]$, with $k_{i-1}<s\leq e\leq k_{i}$, it holds that 
\begin{equation}\label{eqn::exp_rmeans}
    \frac{n_{s,e}}{\sigma^2_n} (\Eeee \bar R_{s,e}-(n+1)/2)^2
    =\frac{n_{s,e}}{\sigma^2_n}\left(\sum_{j=1}^{\ell+1}n_j(p_{ij}-1/2) \right)^2.
\end{equation}
Let $\bfr_{i,j}$ be the $j$th largest element of $\bfr_i$. 
Now, for any $i\in [|\bfr_j|+1]$ and $j\in \{1,2\}$, let $i_j'=\argmin_{m\in [\ell+1],m\geq i} k_m-i$, that is, $i_j'$ is the index of the changepoint immediately to the right of observation $r_{j,i}$. 
By definition, and, in view of \eqref{eqn::exp_rmeans}, it holds that
\begin{align*}
    f(\bfr_1)&=\sigma^{-2}_n\sum_{i=1}^{|\bfr_1|+1}(r_{1,i}-r_{1,i-1})\left(\sum_{j=1}^{\ell+1}n_j(p_{i_1'j}-1/2) \right)^2\\
        &=\sigma^{-2}_n\sum_{i=1}^{|\bfr_2|+1}(r_{2,i}-r_{2,i-1})\left(\sum_{j=1}^{\ell+1}n_j(p_{i_2'j}-1/2) \right)^2=f(\bfr_2). 
\end{align*}
Now, using the above identity, in light of \eqref{eqn::step-1-event-1}, it suffices to show that $f(\bfk)- f(\bfr_1)-c\ell\kappa_n\to \infty$ as $n\to\infty$.

The fact that $f(\bfk)-f(\bfr_1)\geq \Delta_n \tau_{n}^2/4$ follows from the argument in the proof of Lemma 7 of \citep{RAMSAY2023}. We copy it below for the reader:
Let $b_{n,i}= \sum_{m=1}^{\ell+1}n_m (p_{im}-1/2)/n,$ and let $a_n=\Delta_n/2$ and, for any $i\in [n]$, let $i'=\argmin_{m\in [\ell+1],m\geq i} k_m-i$, that is, $i'$ is the index of the changepoint immediately to the right of observation $i$. 
\begin{align*}
    f(\bfk)-f(\bfr_1)&\gtrsim (k_{i^*}-k_{i^*-1}) b_{n,i^*}^2+(k_{i^*+1}-k_{i^*})b_{n,i^*+1}^2 -(k_{i^*}-a_n-k_{i^*-1})b_{n,i^*}^2\\
    &\indent-\frac{\Delta_n}{n^2}\left(\frac{1}{\Delta_n}\sum_{j=k_{i^*}-a_n}^{k_{i^*}+a_n}  \sum_{m=1}^{\ell+1}n_m (p_{j'm}-1/2)  \right)^2-(k_{i^*+1}-k_{i^*}-a_n)b_{n,i^*+1}^2\Bigg)\\
    &= a_n b_{n,i^*}^2+a_nb_{n,i^*+1}^2-\frac{\Delta_n}{n^2}\left(\frac{1}{\Delta_n}\sum_{j=k_{i^*}-a_n}^{k_{i^*}+a_n}  \sum_{m=1}^{\ell+1}n_m (p_{j'm}-1/2)  \right)^2\\
    &= a_n b_{n,i^*}^2/2+a_nb_{n,i^*+1}^2/2-a_n b_{n,i^*}b_{n,i^*+1}\\
    &=a_n(b_{n,i^*}-b_{n,i^*+1})^2/2\\
    &\geq \Delta_n \tau_{n}^2/4.
\end{align*}
We have now shown that there exists universal constants $c,c'>0$ such that $\cW(\bfk)- \cW(\hat\bfk)\gtrsim c'\Delta_n \tau_{n}^2-c\ell\kappa_n.$ 
Therefore, if $\kappa_n\leq c'\Delta_n \tau_{n}^2/\ell$ for all $n\geq 1$, then this yields a contradiction. 

Next, assume that $\hat{\ell}>\ell$. 
First, given $\hat\bfk\subset \hat\bfk\cup \bfk$, we have that $f(\hat\bfk)\leq f(\hat\bfk\cup \bfk)$. 
Using this fact and the assumption that $\hat{\ell}>\ell$, on $A_1(C)^c\cap A_2(C)^c$, we have
\begin{equation*}
    \widehat\cW(\bfk)-\widehat\cW(\hat\bfk)\geq f(\bfk)- f(\hat\bfk)-2C\hat\ell\lambda_{n,\cF,K}+(\hat\ell-\ell)\kappa_n \geq f(\bfk)- f(\hat\bfk\cup \bfk)-2C\hat\ell\lambda_{n,\cF,K}+(\hat\ell-\ell)\kappa_n.
\end{equation*}
Next, similar to the analysis of $f(\bfr_1)$ and $f(\bfr_2)$, we have that $f(\bfk)=f(\hat\bfk\cup \bfk)$. 
Therefore, it holds that 
\begin{equation*}
    \widehat\cW(\bfk)-\widehat\cW(\hat\bfk)\geq (\hat\ell-\ell)\kappa_n-2C\hat\ell\lambda_{n,\cF,K}\geq\hat\ell(\kappa_n-2C\lambda_{n,\cF,K})-\ell\kappa_n.
\end{equation*}
The assumed bounds on $\kappa_n$ yield that  $\hat\ell(\kappa_n-2C\lambda_{n,\cF,K})-\ell\kappa_n\to\infty$ as $n\to\infty$, giving a contradiction. 

Finally, assume that for some $c>0$ there exists $k_{i^*}\in \bfk$ such that $\min_{k\in\bfk}|\hat{k}-k_{i^*}| >c \lambda_{n,\cF,K}/\tau_{n}^2$. 
Define $\bfr'_1$ in the same way as $\bfr_1$ but replace $\Delta_n$ with $c\lambda_{n,\cF,K}/\tau_{n}^2$: 
$$\bfr'_1=\{k_{i^*}-c\lambda_{n,\cF,K}/2\tau_{n}^2,k_{i^*}+c\lambda_{n,\cF,K}/2\tau_{n}^2\}\cup \bfk\backslash k_{i^*}\qquad \text{and}\qquad \bfr'_2=\bfr'_1\cup \hat\bfk.$$ 
Similar to the analysis of $\{\hat{\ell}<\ell\}$, we can write 
\begin{align*}
 \cW(\bfk)- \cW(\hat\bfk)= f(\bfk)- f(\hat\bfk)-2C\ell\lambda_{n,\cF,K}\gtrsim f(\bfk)- f(\bfr_1')-2C\ell\lambda_{n,\cF,K}\geq c\lambda_{n,\cF,K}/4-2C\ell\lambda_{n,\cF,K}.
\end{align*}
Therefore, taking $c$ large enough yields that $c\lambda_{n,\cF,K}/4-2C\ell\lambda_{n,\cF,K}>0$, which yields a contradiction. 
\end{proof}
\begin{proof}[Proof of Theorem \ref{thm::null}]
First, directly from \citep{Chenouri2020DD}, see also \citep{billing}, we have that 
\begin{equation}\label{eqn::bb}
    \widehat{Z}_n(t)=\frac{1}{\sqrt{n}} \sum_{i=1}^{\floor{tn}} \frac{\widehat{R}_{i}-(n+1) / 2}{\sqrt{\left(n^{2}-1\right) / 12}}\cond B(t).
\end{equation} 
In addition, for any $1<k<n$, we have that 
\begin{align*}
\frac{1}{n-k}\sum_{i=k+1}^{n}\widehat{R}_i-\frac{n+1}{2}&=\frac{1}{n-k}\sum_{i=1}^{n}\widehat{R}_i-\frac{1}{n-k}\sum_{i=1}^{k}\widehat{R}_i-\frac{n+1}{2}\\
&=\frac{n}{n-k}\frac{n+1}{2}-\frac{1}{n-k}\sum_{i=1}^{k}\widehat{R}_i-\frac{n+1}{2}\\
&=\left(\frac{n}{n-k}-1\right)\frac{n+1}{2}-\frac{k}{n-k}\frac{1}{k}\sum_{i=1}^{k}\widehat{R}_i\\
&=\frac{k}{n-k}\left(\frac{n+1}{2}-\frac{1}{k}\sum_{i=1}^{k}\widehat{R}_i\right).
\end{align*}
Now, let $\cW'(t)=\cW(\{\floor{tn}\})$. 
Now, observe that for any $t\in(0,1)$, we have that
\begin{align*}\label{eqn::kw_amoc}
\cW'(t)=\cW(\{\floor{tn}\})&=\sigma^{-2}_n\left[\floor{tn}\left(\frac{1}{\floor{tn}}\sum_{i=1}^{\floor{tn}}\widehat{R}_i-\frac{n+1}{2}\right)^2+(n-\floor{tn})\left(\frac{1}{n-\floor{tn}}\sum_{i=\floor{tn}+1}^{n}\widehat{R}_i-\frac{n+1}{2}\right)^2\right]\\
&=\sigma^{-2}_n\left[\floor{tn}\left(\frac{1}{\floor{tn}}\sum_{i=1}^{\floor{tn}}\widehat{R}_i-\frac{n+1}{2}\right)^2+(n-\floor{tn})\left(\frac{\floor{tn}}{n-\floor{tn}}\right)^2\left(\frac{1}{\floor{tn}}\sum_{i=1}^{\floor{tn}}\widehat{R}_i-\frac{n+1}{2}\right)^2\right]\\
&=\sigma^{-2}_n\left[\floor{tn}+(n-\floor{tn})\left(\frac{\floor{tn}}{n-\floor{tn}}\right)^2\right]\left(\frac{1}{\floor{tn}}\sum_{i=1}^{\floor{tn}}\widehat{R}_i-\frac{n+1}{2}\right)^2\\
&=\sigma^{-2}_n\floor{tn}\left[1+\frac{\floor{tn}}{n-\floor{tn}}\right]\left(\frac{1}{\floor{tn}}\sum_{i=1}^{\floor{tn}}\widehat{R}_i-\frac{n+1}{2}\right)^2\\
&=\sigma^{-2}_n\frac{\floor{tn}}{n}\left[\frac{n}{n-\floor{tn}}\right]\left(\sqrt{n}\left(\frac{1}{\floor{tn}}\sum_{i=1}^{\floor{tn}}\widehat{R}_i-\frac{n+1}{2}\right)\right)^2
\end{align*}
It follows from continuous mapping theorem and Slutsky's theorem that 
$$\sup_{1<r\leq n}\cW(r)=\sup_{0<t\leq  1}\cW'(t)\cond \sup_{0<t<1} t(1-t)|B(t)|^2.$$

We now prove the second claim in the theorem. 
Note that
\begin{equation}\label{eqn::ep_kw}
(\hat{k}_1,\hat{k}_2)= \argmax_{r_1,r_2}\frac{12}{n(n+1)}  \left(\left( \sum_{\substack{1\leq i< r_1\\ r_2\leq i\leq n}}^n \frac{\widehat{R}_{i}}{\sqrt{n-r_{2}+r_{1}}} \right)^{2}+\left( \sum_{i= r_1}^{r_2-1} \frac{\widehat{R}_{i}}{\sqrt{r_{2}-r_{1}}} \right)^{2}\right)-3(n+1).
\end{equation}
Let $q_n=\frac{(n^2-1)}{n(n+1)}$, clearly $q_n\rightarrow 1$ as $n\rightarrow\infty$. 
We write $\cW$ as a function of the partial sums of similar form of that of $\widehat{Z}_n(t)$. 
Consider the first term in \eqref{eqn::ep_kw}. We have that
\begin{align*}
   \left(\frac{12}{n(n+1)}\right)^{1/2} \sum_{1 \leq j<k_{1} \atop k_{2} \leq j \leq n}^{n} \frac{\widehat{R}_{j}-(n+1)/2}{\sqrt{n-k_{2}+k_{1}}}&=\left(\frac{q_n}{1-k_{2}/n+k_{1}/n}\right)^{1/2}\frac{1}{\sqrt{n}}\left(\sum_{j=1 }^{k_{1}-1} \frac{\widehat{R}_{j}-(n+1)/2}{(n^2-1)/12}+\sum_{j=k_2 }^{n} \frac{\widehat{R}_{j}-(n+1)/2}{(n^2-1)/12}\right)\\
    &=\left(\frac{q_n}{1-k_{2}/n+k_{1}/n}\right)^{1/2}\frac{1}{\sqrt{n}}\left(\sum_{j=1 }^{k_{1}-1} \frac{\widehat{R}_{j}-(n+1)/2}{(n^2-1)/12}-\sum_{j=1 }^{k_2-1} \frac{\widehat{R}_{j}-(n+1)/2}{(n^2-1)/12}\right)\\
    &=-\left(\frac{q_n}{1-k_{2}/n+k_{1}/n}\right)^{1/2}\frac{1}{\sqrt{n}}\left(\sum_{j=k_1 }^{k_{2}-1} \frac{\widehat{R}_{j}-(n+1)/2}{(n^2-1)/12}\right). 
\end{align*}
For the second term in \eqref{eqn::ep_kw}, we have that 
\begin{align*}
    \left(\frac{12}{n(n+1)}\right)^{1/2} \sum_{j=k_{1} }^{k_2-1} \frac{\widehat{R}_{j}-(n+1)/2}{\sqrt{k_{2}-k_{1}}}&= \left(\frac{q_n}{k_{2}/n-k_{1}/n}\right)^{1/2}\frac{1}{\sqrt{n}}\left(\sum_{j=k_1 }^{k_{2}-1} \frac{\widehat{R}_{j}-(n+1)/2}{(n^2-1)/12}\right).
\end{align*}
So, it follows that 
$$\cW(k_1,k_2)=\frac{q_n}{(k_{2}/n-k_{1}/n)(1-k_{2}/n+k_{1}/n)}\left(\frac{1}{\sqrt{n}}\sum_{j=k_1 }^{k_{2}-1} \frac{\widehat{R}_{j}-(n+1)/2}{(n^2-1)/12}\right)^2.$$
We can write $\cW$ as a function of $0\leq t_1\leq t_2\leq 1$ as $\cW(t_1,t_2)$, where $k_1,k_2$ are replaced with $\floor{t_1 n},\ \floor{t_2 n}$ respectively. 
We can then write 
$$\cW(t_1,t_2)\coloneqq g_n(t_1,t_2)\cW'(t_1,t_2),$$
where $$g_n(t_1,t_2)=\frac{1}{(t_2-t_1)(1-t_2+t_1)}+o(1)  \qquad\text{ and }\qquad \cW'(t_1,t_2)=\left(\frac{1}{\sqrt{n}}\left(\sum_{j=\floor{nt_1} }^{\floor{nt_2}-1} \frac{\widehat{R}_{j}-(n+1)/2}{(n^2-1)/12}\right)\right)^2.$$

Recall that $\widehat{Z}_n(t)\cond B(t)$, and it is also clear that $\cW'(t_1,t_2)=(\widehat{Z}_n(t_1)-\widehat{Z}_n(t_2))^2$. 
$\cW'(t_1,t_2)$ is a continuous functional of $\widehat{Z}_n(t)$, and so continuous mapping theorem gives that
$$\cW'(t_1,t_2)=\left(\frac{1}{\sqrt{n}}\left(\sum_{j=\floor{nt_1} }^{\floor{nt_2}-1} \frac{\widehat{R}_{j}-(n+1)/2}{(n^2-1)/12}\right)\right)^2\cond (B(t_2)-B(t_1))^2.$$
and 
\begin{align*}
    \sup_{t_1<t_2}\cW((\floor{nt_1},\floor{nt_2}))\cond \sup_{t_1<t_2}  \left(\frac{1}{(t_{2}-t_{1})(1-t_{2}+t_{1})}\right)(B(t_2)-B(t_1))^2.
\end{align*}
\end{proof}
\section{Additional Simulation Results}\label{app::sim}
Below we have some additional simulation results for the epidemic change, the effects of the derivatives and the effects of dependency in the data. 
\begin{table}[ht]
\caption{Table of empirical powers for the epidemic FKWC test when there was an epidemic-type magnitude change, an epidemic-type shape change and no change.}
\centering
\begin{tabular}{rrrrrrrrrr}
 \hline
 change type &\multicolumn{9}{c}{Scale ($\beta$)}\cr
  \hline
dist. & \multicolumn{3}{c}{$\cG$} & \multicolumn{3}{c}{$t_3$}&\multicolumn{3}{c}{$\mathcal{SG}$}\\ 
  \hline
$n$ & 100 & 200 & 500 & 100 & 200 & 500 & 100 & 200 & 500 \\ 
  \hline
  $\MFHD$& 0.92 & 1.00 & 1.00 & 0.14 & 0.49 & 0.98 & 0.88 & 1.00 & 1.00\\ 
$\RP$  & 0.83 & 0.99 & 1.00 & 0.14 & 0.48 & 0.96 & 0.86 & 1.00 & 1.00 \\ 
 \hline
 change type &\multicolumn{9}{c}{Shape ($\alpha$)}\cr
  \hline
dist. & \multicolumn{3}{c}{$\cG$} & \multicolumn{3}{c}{$t_3$}&\multicolumn{3}{c}{$\mathcal{SG}$}\\ 
  \hline
$n$ & 100 & 200 & 500 & 100 & 200 & 500 & 100 & 200 & 500 \\ 
  \hline
  $\MFHD$ & 0.73 & 0.96 & 1.00 & 0.07 & 0.37 & 0.92 & 0.70 & 0.98 & 1.00\\ 
$\RP$ & 0.76 & 0.96 & 1.00 & 0.13 & 0.49 & 0.95 & 0.76 & 0.97 & 1.00 \\ 
 \hline
 change type &\multicolumn{9}{c}{No Change}\cr
  \hline
dist. & \multicolumn{3}{c}{$\cG$} & \multicolumn{3}{c}{$t_3$}&\multicolumn{3}{c}{$\mathcal{SG}$}\\ 
  \hline
$n$ & 100 & 200 & 500 & 100 & 200 & 500 & 100 & 200 & 500 \\ 
  \hline
  $\MFHD$ & 0.00 & 0.00 & 0.01 & 0.00 & 0.00 & 0.02 & 0.00 & 0.00 & 0.01 \\ 
$\RP$ & 0.00 & 0.00 & 0.02 & 0.00 & 0.00 & 0.02 & 0.00 & 0.00 & 0.00 \\ 
   \hline
\end{tabular}
\end{table}
\begin{figure}[t!]
\begin{minipage}[c]{.45\textwidth} 
    \centering
    \includegraphics[width=1\textwidth]{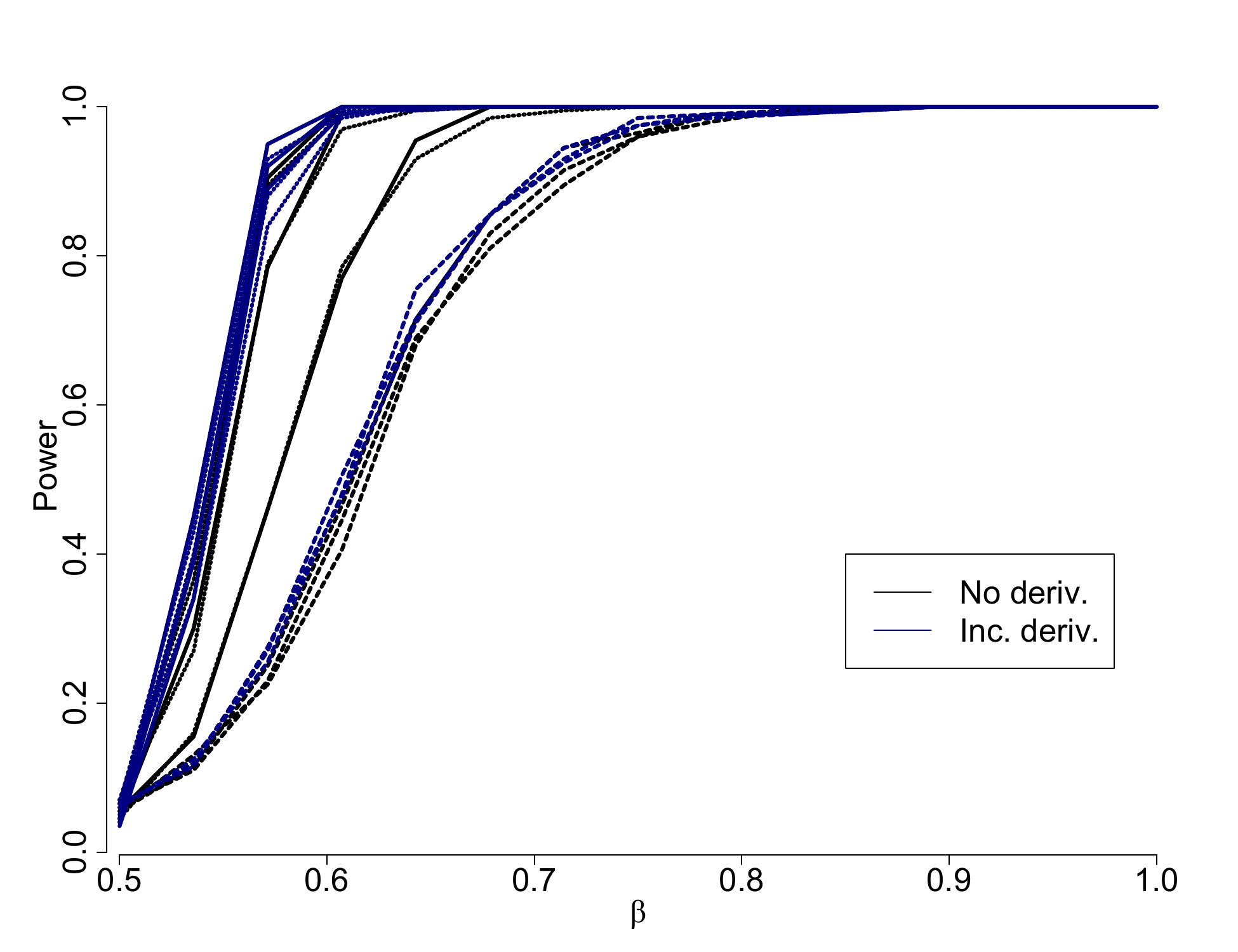}
     \caption*{(a) }
\end{minipage}
\begin{minipage}[c]{.45\textwidth} 
    \centering
  \includegraphics[width=1\textwidth]{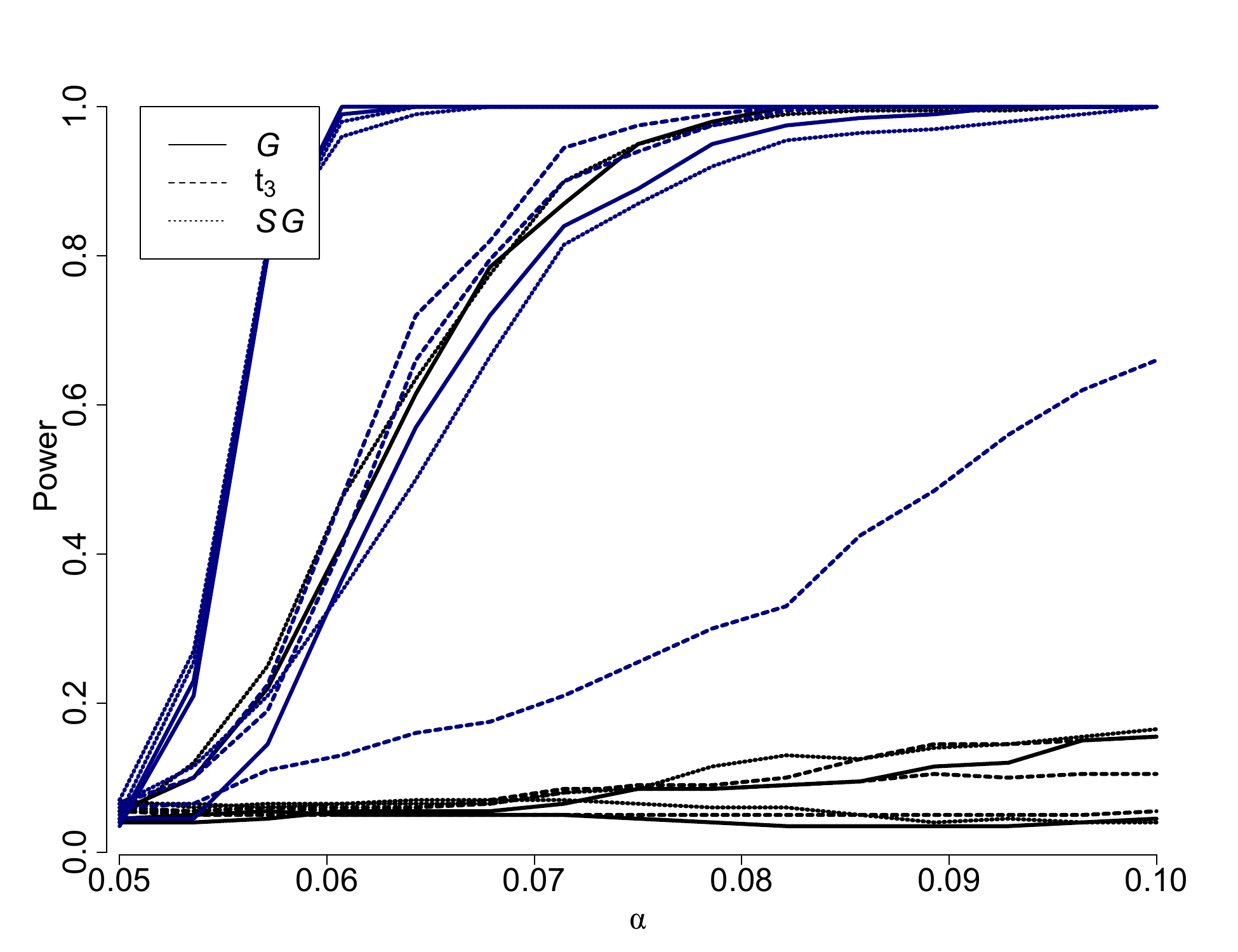}
    \caption*{(b)  }
\end{minipage}
 \caption{Comparison of the FKWC methods based on the depth value sequence with the derivatives to the FKWC methods based on the depth value sequence without the derivatives under (a) scale changes and (b) shape changes.    }
\end{figure}

\begin{table}[ht]
\caption{Table of empirical powers at the $5\%$ level of significance for the AMOC FKWC test under the functional autoregressive model discussed in the simulation study of \citet{Sharipov2019}. The highest power reported for the test of \citet{Sharipov2019} was 0.925, \citep[][see Table 4]{Sharipov2019}. }
\centering
\begin{tabular}{rrrrrrr}
  \hline
 $(d_1,d_2)$ & (0,0) & (0.4,0) & (0.8,0) & (0,0.4) & (0,0.8) & (0.4,0.4) \\ 
  \hline
 $\MFHD$ & 0.10 & 0.98 & 1.00 & 0.98 & 1.00 & 1.00 \\ 
$\RP$ & 0.06 & 0.97 & 1.00 & 0.97 & 1.00 & 1.00 \\ 
   \hline
\end{tabular}
\label{tab::shar_ar}
\end{table}

\end{document}